\RequirePackage{fixltx2e}
\documentclass[12pt,canadian,twocolumn]{svjour3}
\usepackage[T1]{fontenc}
\usepackage[latin9]{inputenc}
\usepackage{geometry}
\PassOptionsToPackage{inline,nomargin}{fixme}
\usepackage{babel}
\usepackage{url}
\usepackage{enumitem}
\usepackage{fixme}
\usepackage{amsmath}
\usepackage{amssymb}
\usepackage{graphicx}
\usepackage{esint}
\usepackage[unicode=true,
 bookmarks=true,bookmarksnumbered=false,bookmarksopen=false,
 breaklinks=false,pdfborder={0 0 1},backref=false,colorlinks=false]
 {hyperref}

\makeatletter

\providecommand{\tabularnewline}{\\}
\newcommand{\lyxdot}{.}

\usepackage{siunitx}
\usepackage{listings}
\usepackage{booktabs}
\usepackage{tabto}
\usepackage[svgnames]{xcolor}
\usepackage{pifont}

\usepackage{cite}

\definecolor{marginsymbolred}{HTML}{CD5C5C}
\definecolor{marginsymbolblue}{HTML}{0000FF}

\fxsetup{draft,theme=color}

\usepackage{tabulary}
\usepackage{multirow}
\usepackage{listings}
\usepackage{color}
 
\definecolor{codegreen}{rgb}{0,0.6,0}
\definecolor{codegray}{rgb}{0.5,0.5,0.5}
\definecolor{codepurple}{rgb}{0.58,0,0.82}
\definecolor{backcolour}{rgb}{0.95,0.95,0.92}
 
\lstdefinestyle{mystyle}{
    backgroundcolor=\color{backcolour},   
    commentstyle=\color{codegreen},
    keywordstyle=\color{magenta},
    numberstyle=\tiny\color{codegray},
    stringstyle=\color{codepurple},
    basicstyle=\footnotesize\ttfamily,
    breakatwhitespace=false,         
    breaklines=true,                 
    captionpos=b,                    
    keepspaces=true,                 
    numbersep=5pt,                  
    showspaces=false,                
    showstringspaces=false,
    showtabs=false,                  
    tabsize=2
}
 
\makeatother

\begin{document}
\inputencoding{latin9}\global\long\def\sbn#1#2{\{#1\:|\:#2\}}%

\global\long\def\norm#1{\Vert#1\Vert}%

\global\long\def\abs#1{|#1|}%
\global\long\def\vabs#1{\left|#1\right|}%

\global\long\def\xhat#1{\hat{#1}}%
\global\long\def\rsph{r}%
\global\long\def\rcyl{\rho}%

\global\long\def\xx{\xhat x}%
\global\long\def\xy{\xhat y}%
\global\long\def\xz{\xhat z}%

\global\long\def\d{\mathrm{d}}%

\global\long\def\f#1#2#3{#1:\ #2\rightarrow#3}%

\global\long\def\k#1{\mathop{\mathrm{#1}}\nolimits}%

\global\long\def\braket#1#2{\Braket{#1|#2}}%

\global\long\def\bra#1{\Bra{#1}}%

\global\long\def\ket#1{\Ket{#1}}%

\global\long\def\C#1{\cancel{#1}}%

\global\long\def\BC#1{\bcancel{#1}}%

\global\long\def\u#1{\mathrm{\:#1}}%

\global\long\def\li{\k{li}}%

\title{Simudo: a device model for intermediate band materials}
\author{Eduard C. Dumitrescu$^{1}$ \and Matthew M. Wilkins$^{1,2}$ \and
Jacob J. Krich$^{1,2}$}
\institute{J.~J.~Krich \\jkrich@uottawa.ca \\$^{1}$Department of Physics,
University of Ottawa, ON, Canada \\$^{2}$School of Electrical Engineering
and Computer Science, University of Ottawa, ON, Canada}
\date{Received: date / Accepted: date}
\maketitle
\begin{abstract}
We describe Simudo, a free Poisson/drift-diffusion steady state device
model for semiconductor and intermediate band materials, including
self-consistent optical absorption and generation. Simudo is the first
freely available device model that can treat intermediate band materials.
Simudo uses the finite element method (FEM) to solve the coupled nonlinear
partial differential equations in two dimensions, which is different
from the standard choice of the finite volume method in essentially
all commercial semiconductor device models. We present the continuous
equations that Simudo solves, show the FEM formulations we have developed,
and demonstrate how they allow robust convergence with double-precision
floating point arithmetic. With a benchmark semiconductor pn-junction
device, we show that Simudo has a higher rate of convergence than
Synopsys Sentaurus, converging to high accuracy with a considerably
smaller mesh. Simudo includes many semiconductor phenomena and parameters
and is designed for extensibility by the user to include many physical
processes.
\end{abstract}

\section{Introduction}

Device models are essential components of the development of semiconductor
devices, from transistors to solar cells to lasers. Standard semiconductor
device models, such as Synopsys Sentaurus, treat materials with 0,
1, or 2 bands (i.e., dielectrics, metals, and semiconductors, respectively)
along with an electrostatic potential. At a given location in a given
material, each band has its own carrier concentration, with particle
motion given by diffusion and electric-field-induced drift. Since
the electric field itself depends on particle motion, the resulting
Poisson/drift-diffusion (PDD) equations are nonlinear and require
numerical solution in the general case \cite{Bank1983,Fichtner1983a,Markowich1986,Piprek18,Schenk1998}.

\begin{figure}
\includegraphics[width=1\columnwidth]{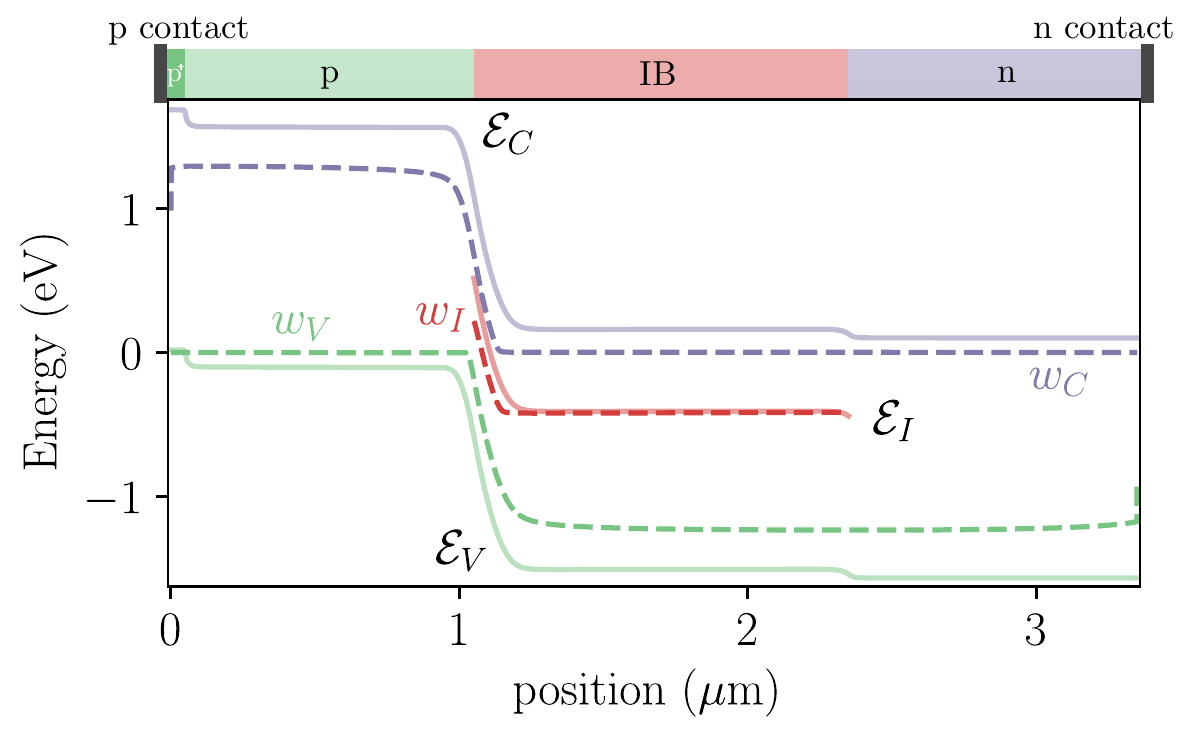}

\caption{\label{fig:IB_intro_bands}Band diagram of illuminated 1D intermediate
band device at short circuit. Device structure of p-IB-n device shown
at top, and the IB material has an extra band contained entirely inside
the semiconductor band gap. Solid lines show the band edge energies
$\mathcal{E}_{C}$, $\mathcal{E}_{V}$ and the IB energy $\mathcal{E}_{I}$.
Dashed lines show the quasi-Fermi levels $w_{C}$, $w_{V}$, and $w_{I}$.
Parameters as in Sec.~\ref{subsec:Marti_comparison}.}
\end{figure}

A new class of materials, called intermediate band (IB) materials,
has been developed over the last 20 years with the goal of improving
solar cell efficiency and producing effective infrared photodetectors
\cite{Luque97,Okada15,Mailoa14,Berencen17,Wang18}. These IB materials
are like semiconductors except they have an extra band of allowed
electronic energy levels above the valence band (VB) and below the
conduction band (CB), as shown in Fig.~\ref{fig:IB_intro_bands}.
Such a band structure permits optical absorption from VB to IB and
from IB to CB, which is the key to the increased solar cell efficiency
\cite{Luque97}. It is also possible to consider multiple IBs, though
such materials have not yet been realized in practice \cite{Brown02}.

Where IB devices have been made, they have not generally been highly
efficient, which is believed to be largely due to  fast nonradiative
recombination processes \cite{Okada15,Marti06,Wang09,Lopez11,Mailoa14,Berencen17,Wang18,Sullivan15}.
It has not been possible, however, to perform standard device modeling
to optimize these devices, to determine the ideal layer thicknesses,
doping levels, etc.,~since standard semiconductor device models do
not allow the possibility of treating a third band. Therefore, we
do not know what efficiencies existing IB materials could permit,
if they were optimized. Interpreting experiments on IB materials and
designing the best devices require device modeling capabilities.

In order to describe the basic physics of IB devices, one must be
able to describe
\begin{enumerate}
\item Optical processes between CB, VB, and IB, with rates dependent on
IB filling fraction $f$,
\item Nonradiative processes between CB, VB, and IB, with rates dependent
on $f$,
\item Carrier transport within the IB,
\item Junctions with standard semiconductors.
\end{enumerate}
There is a large array of standard numerical semiconductor device
models based on the coupled Poisson and carrier-continuity equations,
including general purpose ones, such as Synopsys Sentaurus and Silvaco,
as well as more specialized models such as Crosslight, which includes
modelling of quantum well physics and a coupled treatment of carrier-density
dependent optics for lasers, and TiberCAD \cite{Maur08}. Nextnano++
also includes features specific to quantum structures and can solve
an 8-band k.p model self-consistently with the Poisson and drift-diffusion
calculations \cite{Birner2007}. There are also more focused ones,
such as PC1D \cite{Clugston97,Haug16}, AFORS-HET\cite{Varache15},
SCAPS \cite{Burgelman00}, and Solcore \cite{Alonso-Alvarez2018},
which are 1D models focused on solar cells. Many of these models allow
treatment of deep-lying states inside the semiconductor band gap,
primarily as Shockley-Read-Hall (SRH) trapping and recombination centers
\cite{Shockley52}. Sentaurus and PC1D, for example, do not permit
optical generation from the deep-lying states. SCAPS does permit both
thermal and optical processes, but does not consider transport of
carriers inside the defect band. We do not attempt a full characterization
of all the available device models, but Table \ref{tab:devicemodels}
shows which of these requirements are met by these device models.

\begin{table}
\caption{Comparison of selected device models \label{tab:devicemodels}}

\setlength{\tabcolsep}{1pt}
\providecommand{\tabY}{\centering Y}
\providecommand{\tabN}{\centering N}
\begin{tabular}{p{2.1cm}p{1.8cm}p{2.3cm}p{1.1cm}p{1.4cm}p{0.5cm}}
\toprule   & \centering IB \mbox{optics} with \mbox{photofilling} & \centering IB \mbox{nonradiative} \mbox{processes} & \centering IB \mbox{transport} & \centering \mbox{Junctions} & \centering 2D\tabularnewline 
\midrule 
Sentaurus & \tabN & \tabY & \centering limited & \tabY & \tabY\tabularnewline 
PC1D \cite{Clugston97} 
& \tabN & \tabY & \tabN & \tabY & \tabN\tabularnewline 
SCAPS \cite{Burgelman00} 
& \tabY & \tabY & \tabN & \tabY & \tabN\tabularnewline 
TiberCAD \cite{Maur08}
& \tabN & \tabY & \tabY & \tabY & \tabY \tabularnewline
\midrule  
Mart\'i \cite{Marti02} 
& \tabN & \tabN & \tabY & \tabN & \tabN\tabularnewline
\mbox{Strandberg} \cite{Strandberg11} 
& \centering \tabY & \tabN & \tabY & \tabN & \tabN\tabularnewline 
Tobias \cite{Tobias11}, Yoshida  \cite{Yoshida12a} 
& \tabY & \tabN & \tabY & \tabY & \tabN\tabularnewline 
\midrule  
Simudo & \tabY & \tabY & \tabY & \tabY & \tabY\tabularnewline \bottomrule
\end{tabular}

\end{table}
There have been a number of device models developed specifically for
IB materials, mostly for solar cells, all in steady state. These include
traditional \cite{Luque97,Cuadra04a,Levy08,Hu10}and Boltzmann-approximation
\cite{Strandberg17,Strandberg17a} detailed balance models, semianalytic
models in the drift \cite{Lin09} and diffusion \cite{Marti02,Navruz08,Krich14}
limits, and PDD models \cite{Yoshida10,Strandberg11,Tobias11,Yoshida12a}.
The semianalytic models are specific to either the drift or diffusive
limits, while the PDD models allow treatment of IB regions that are
neither fully depleted nor fully quasi-neutral. A comparison of the
features of the PDD models is also included in Table~\ref{tab:devicemodels}.
To our knowledge, none has been released as open-source software.

Here we introduce Simudo, a free and open source steady state PDD
solver with self-consistent optics for arbitrary numbers of bands.
Simudo uses the finite element method (FEM) to solve the coupled Poisson,
drift-diffusion, and Beer-Lambert optical propagation equations self-consistently,
when necessary including changing $f$ according to local generation
and recombination, with associated changes in the optical absorption
coefficient. Simudo has built-in radiative recombination, Shockley-Read
trapping, and SRH recombination models in the non-degenerate limit
and is straightforward to extend to include other models of generation
or recombination. All of the band parameters, from energies to mobilities
to cross sections, can vary in space or as functions of other parameters.

Simudo has a number of innovations in its formulation of the problem,
described below, and allows high-accuracy simulation of benchmark
semiconductor problems while working with 64-bit arithmetic, making
it useful both for standard semiconductor simulations as well as for
IB devices. It is written in the Python programming language, using
the FEniCS platform to solve the FEM problem \cite{AlnaesBlechta2015a}.
It exposes an easy-to-use API for defining problems and extracting
results. It is designed for two-dimensional systems and is available
for download at \href{https://github.com/simudo/simudo}{https://github.com/simudo/simudo}.

Semiconductor device modeling is typically performed using the finite
volume method (FVM), which ensures local charge conservation at each
cell in the domain \cite{Eymard00}. Finite element methods are less
common in semiconductor device models, though there are a number of
examples \cite{He91,Maur08,Nachaoui99,Bochev15}. FEM is widely used
for related advection-diffusion problems in computational fluid dynamics
(CFD) studies \cite{Cockburn00}. Commercial packages for CFD, as
for semiconductor device modeling, are generally based on the FVM
method. TiberCAD, a commercial device modeling package with many novel
features, uses FEM with continuous basis functions\cite{Maur08}.
Finite element methods including discontinuous local basis functions,
called discontinuous Galerkin (DG) methods, also permit local charge
conservation \cite{Cockburn00}, and they have recently begun to be
applied to semiconductor device problems \cite{Kumar16,Kumar17}.
FEM methods simplify consideration of complicated simulation domains
and in theory allow higher-order convergence of solutions, but performance
of such methods can only be determined with testing. We use such a
DG-FEM method here to produce a general purpose steady state PDD solver
capable of treating IB systems, and we show that Simudo realizes the
higher-order convergence with mesh size, converging much more rapidly
than Synopsys Sentaurus as the mesh spacing is reduced. As shown in
Sec.~\ref{subsec:num-transport-sentaurus}, for a reference pn-diode,
Simudo demonstrates quartic self-convergence with mesh density while
FVM-based Synopsys Sentaurus demonstrates only quadratic convergence.
In the reference problem, Simudo achieves 5-6 digits of convergence
with 193 mesh points while Sentaurus requires more than 3000. Simudo
provides both a flexible framework for the study of IB devices and
also a freely available example of a DG-FEM semiconductor device model.

In Section~\ref{sec:Statement-of-problem}, we define the coupled
partial differential equations (PDEs) Simudo solves. Section \ref{sec:numerical-method}
describes the heart of Simudo, giving in detail the conversion of
the equations of Section~\ref{sec:Statement-of-problem} to the weak
forms solved using FEM. This section describes the choices for dynamical
variables, the weak forms used for FEM, and how these choices enable
Simudo to achieve accuracy despite the problems of finite precision
arithmetic. This section concludes with a comparison to Synopsys Sentaurus
on a benchmark pn-diode, showing the high quality of Simudo's results.
Section~\ref{subsec:PN-junction-example} gives examples of setting
up a simple problem using the API, including examples of its convenient
topology definitions and Section \ref{subsec:Auger_recombination}
demonstrates the extensibility of Simudo to include new physical processes
(in this case, Auger recombination). Section \ref{subsec:Marti_comparison}
demonstrates the use of Simudo to analyze a system originally studied
in Ref.~\citenum{Marti02}, showing that its model works better than
had been anticipated in the case with equal subgap optical absorption
cross sections, but that unequal subgap absorption cross sections
produce more complicated phenomena that require IB transport to describe
properly.

\section{Statement of problem\label{sec:Statement-of-problem}}

In this section, we describe the mathematical model of the steady
state PDD and optical problems we use in Simudo. Carriers both drift
in response to electric fields and diffuse. Carriers are generated
optically and recombine using a variety of models. The local carrier
concentration determines both the electric field and the optical absorption
coefficients, so the transport, Poisson, and optical propagation equations
are all coupled. Symbols used in this manuscript are summarized in
Table \ref{tab:symbols}.

\begin{table}[h]
\noindent \begin{centering}
\caption{\label{tab:symbols}Common symbols used in this manuscript.}
\par\end{centering}
\begin{tabular}{c p{7.5cm}} 
\toprule  \textbf{Symbol} & \textbf{Definition}\tabularnewline 
\midrule 
$u_{k}$ & Carrier density in band $k$\tabularnewline
$\mathcal{E}_{k}$ & Band edge energy of band $k$; central energy of IB $k$\tabularnewline
$s_k$	& 	Sign of carriers in band $k$ (+ for VB, - for CB)\tabularnewline
$w_{k}$ & Quasi-Fermi level of carriers in band $k$\tabularnewline
$\vec{j}_{k}$ & Current density in band $k$\tabularnewline
$\mu_{k}$ & Mobility of carriers in band $k$\tabularnewline
$\mathcal{D}_{k}$ & Diffusion coefficient of carriers in band $k$\tabularnewline
$S_{k}$ & Surface recombination velocity of carriers in band $k$\tabularnewline
\multirow{2}{*}{$N_{k}$} & Effective density of states for nondegenerate band $k$\tabularnewline
 & Integrated density of states for intermediate band $k$\tabularnewline
$f_{k}$ & Filling fraction $f_{k}=u_{k}/N_{k}$ of IB $k$\tabularnewline
$f_{k,\pm}$ & $(1-f_k)$  for positive sign and $f_k$ for negative sign \tabularnewline
$f_{k,0}$ & Charge neutral filling fraction of IB $k$\tabularnewline
$g_{k}$ & Net generation in band $k$ due to all generation and recombination processes\tabularnewline
$T$ & Temperature\tabularnewline $k_{B}$ & Boltzmann constant\tabularnewline
$q$ & Elementary charge\tabularnewline
$\phi,\vec{E},\rho$ & Electrostatic potential, electric field, and charge density\tabularnewline
$\alpha_{fi,\lambda}$ & Optical absorption coefficient from band $i$ to $f$ at vacuum wavelength $\lambda$\tabularnewline
$\sigma_{fi}^{\text{opt}}$ & Optical cross section from band $i$ to $f$\tabularnewline
$\Phi_{\lambda,\hat{s}}$ & Photon spectral flux density at wavelength lambda in direction $\hat{s}$\tabularnewline
$\Phi_{[\lambda_{1},\lambda_{2}],\hat{s}}$ & Photon flux in direction $\hat{s}$ from $\lambda_{1}$ to $\lambda_{2}$, i.e., $\int_{\lambda_{1}}^{\lambda_{2}}\Phi_{\lambda,\hat{s}}\:\d\lambda$ \tabularnewline
$\hat{n}$ & Surface normal vector\tabularnewline
$\hat{s}$ & Direction of light propagation\tabularnewline
\bottomrule 
\end{tabular}
\end{table}

\subsection{Carrier transport and generation}

We consider a CB, a VB, and some number of IBs under the assumption
that the carrier population in each band is in local quasi-equilibrium
with a temperature $T$ and quasi-Fermi level $w_{k}$, where $k$
can be one of $\{C,V,I\}$ for the CB, VB, and IB, respectively. In
the case of multiple IBs, $k$ can take values $I_{1},I_{2},\dots$,
indexing the various IBs, but we simplify the following discussion
to consider the case of just one IB, indexed as $I$.

In the most common approximation of semiconductor device modeling,
the carrier dynamics in each band can be described by the drift-diffusion
equation and the continuity equation. Letting $u_{k}$ represent the
carrier concentration in band $k$, $u_{V}$ and $u_{C}$ are the
hole and electron concentrations, respectively, which we use interchangeably
with their standard symbols, $p$ and $n$. We let $s_{k}=\pm1$ give
the charge of the carriers in band $k$, $+1$ for the VB and -1 for
the CB. Then
\begin{subequations}
\label{eq:dd}
\begin{align}
\vec{j}_{k} & =\overbrace{q\mu_{k}u_{k}\vec{E}}^{\text{drift}}-\overbrace{s_{k}q\mathcal{D}_{k}\vec{\nabla}u_{k}}^{\textnormal{diffusion}}\label{eq:dd-j}\\
\frac{\partial u_{k}}{\partial t} & =-s_{k}\frac{1}{q}\vec{\nabla}\cdot\vec{j}_{k}+g_{k},\label{eq:dd-g}
\end{align}
\end{subequations}
where $\vec{j}_{k}$ is the current density of carriers in band $k$,
$\mu_{k}$ is the carrier mobility, $\mathcal{D}_{k}$ is the carrier
diffusion constant, $\vec{E}$ is the electric field, $q$ is the
elementary charge, and $g_{k}$ contains all the generation, trapping,
and recombination processes (see Section~\ref{subsec:g}). For non-degenerate
bands in which $w_{k}$ is sufficiently far from the band edge $\mathcal{E}_{k}-q\phi$,
we can write 
\begin{align}
u_{k} & =N_{k}e^{-s_{k}(w_{k}+q\phi-\mathcal{E}_{k})/k_{B}T},\label{eq:nondeg-u}
\end{align}
where $N_{k}$ is the effective density of states of band $k$, $\phi$
is the electrostatic potential, and $k_{B}$ is Boltzmann's constant.

Then, assuming $\mathcal{E}_{k}$ is spatially constant,
\begin{align}
\vec{\nabla}u_{k} & =-s_{k}N_{k}e^{-s_{k}(w_{k}+q\phi-\mathcal{E}_{k})/k_{B}T}\frac{1}{k_{B}T}\vec{\nabla}(w_{k}+q\phi)\\
 & =-s_{k}\frac{u_{k}}{k_{B}T}\vec{\nabla}(w_{k}+q\phi),
\end{align}

For such nondegenerate bands, the Einstein relation gives $\mu_{k}=q\mathcal{D}_{k}/k_{B}T$,
from which Eq.~\ref{eq:dd-j} gives \cite{Fichtner1983a}
\begin{align}
\vec{j}_{k} & =\mu_{k}u_{k}\,\vec{\nabla}w_{k},\label{eq:j-mu-rel}
\end{align}
which we use instead of Eq.~\ref{eq:dd-j}. Equation~\ref{eq:j-mu-rel}
also applies to the case of degenerate bands, as shown in \cite{Poupaud1991a},
even though the Einstein relation requires a modification. Moreover,
Eq.~\ref{eq:j-mu-rel} applies in the case of spatially-varying band
structure (e.g., spatially-varying $N_{c},\mathcal{E}_{c}$)\cite{Marshak1984a},
so it is considerably more general than this derivation.

Since an intermediate band is often partially filled, we cannot model
it using the non-degenerate approximation of Eq.~\ref{eq:nondeg-u}.
We write $D_{I}(E)$ for the density of states of the IB, such that
$N_{I}=\int\d E\:D_{I}(E)$ is the total density of IB states. If
the IB has quasi-Fermi level $w_{I}$, the electron concentration
is
\begin{align}
u_{I} & =\int\d E\:\frac{D_{I}(E)}{e^{(E-w_{I}-q\phi)/k_{B}T}+1}.
\end{align}

If the bandwidth of the IB is narrow relative to $k_{B}T$, we can
approximate the IB density of states as a Dirac delta $D_{I}(E)=N_{I}\delta(E-\mathcal{E}_{I})$,
and so
\begin{align}
u_{I} & =N_{I}\,\underbrace{\frac{1}{e^{(\mathcal{E}_{I}-w_{I}-q\phi)/k_{B}T}+1}}_{f_{I}},\label{eq:deg-uI}
\end{align}
where $f_{I}$ is the filling fraction of the IB, and can be written
as $f_{I}=f(\mathcal{E}_{I}-w_{I}-q\phi)$ where $f(E)$ is the Fermi
function. We work in this limit for the remainder of this manuscript.
Extending beyond this sharp-IB case is not difficult but requires
more cumbersome notation.

\subsection{\label{subsec:g}Carrier generation and recombination}

Each band's continuity equation (Eq.~\ref{eq:dd-g}) has a generation
term $g_{k}$. This term is the sum of contributions from all generation
and recombination processes to the band, which depend on which physical
models are included in the simulations. We now specify the details
of optical generation $g_{k}^{\text{opt}}$ and a variety of recombination
processes $r_{k}$, each of which enters either as a negative or positive
contribution to $g_{k}$, as required for the process.

\subsubsection{Optical carrier generation}

Modeling optical carrier generation requires modeling the changing
light intensity through the device. We use a simple Beer-Lambert model
for optical propagation and absorption
\begin{align}
\vec{\nabla}\Phi_{\lambda,\hat{s}}\cdot\hat{s} & =-\alpha_{\lambda}\Phi_{\lambda,\hat{s}}\label{eq:beerlambert}
\end{align}
where $\Phi_{\lambda,\hat{s}}$ is the photon spectral flux at vacuum
wavelength $\lambda$ and direction of propagation $\hat{s}$ and
$\alpha_{\lambda}$ is the total absorption coefficient, which can
be written as 
\[
\alpha_{\lambda}=\sum_{i,f}\alpha_{fi,\lambda},
\]
where $\alpha_{fi,\lambda}$ is the absorption coefficient for the
optical process at wavelength $\lambda$ that moves a carrier from
band $i$ to band $f$. In the usual semiconductor case, $\alpha_{VC,\lambda}=0$
and $\alpha_{CV,\lambda}$ is finite for $\lambda$ corresponding
to energies larger than the band gap. Free-carrier absorption is included
in $\alpha_{ii,\lambda}.$ The carrier generation rate in band $k$
due to optical processes is then
\begin{align}
g_{k}^{\textnormal{opt}} & =-s_{k}\int\d\lambda\:\bigg(\sum_{i}\alpha_{ki,\lambda}-\sum_{f}\alpha_{fk,\lambda}\bigg)\,\Phi_{\lambda,\hat{s}}.
\end{align}
Further details of the optical propagation model are described in
Section~\ref{subsec:opt}.

In nondegenerate bands, there are always enough carriers to excite
in or out of a band. That is, the valence band always has electrons
available, and the conduction band has empty states available to be
filled, so the absorption coefficient $\alpha_{CV,\lambda}$ is insensitive
to the free carrier density in the bands. In an IB, however, the VB$\rightarrow$IB
process requires empty states in the IB while the IB$\rightarrow$CB
process requires filled states in the IB. To capture this phenomenon,
we write
\begin{align}
\alpha_{CI,\lambda} & =\sigma_{CI,\lambda}^{\textnormal{opt}}u_{I}\label{eq:alpha_CI}\\
\alpha_{IV,\lambda} & =\sigma_{IV,\lambda}^{\textnormal{opt}}(N_{I}-u_{I})\label{eq:alpha_IV}
\end{align}
 where $\sigma_{fi,\lambda}^{\text{opt}}$ is the optical capture
cross section from band $i$ to $f$ at wavelength $\lambda$ of a
single intermediate state. We can combine these equations into a single
expression,
\begin{equation}
\alpha_{kI,\lambda}=\sigma_{kI,\lambda}^{\text{opt}}u_{I,s_{k}},\label{eq:alpha_kI}
\end{equation}
where $u_{I,-}$ is just $u_{I}$ and $u_{I,+}=N_{I}-u_{I}$ is the
number of holes in band $I$, and $\alpha_{kI}$ is understood to
be $\alpha_{IV}$ when $k=V$. Since $\alpha_{fi,\lambda}$ depends
on the carrier concentrations, and the carrier concentrations depend
on $\alpha_{fi,\lambda}$ (through the generation rate $g_{k}^{\textnormal{opt}}$),
the transport and the optical models feed into each other, so they
must be solved in a self-consistent manner.

\subsubsection{Recombination and trapping\label{subsec:Recombination-and-trapping}}

Simudo offers several built-in radiative and nonradiative recombination
and trapping mechanisms using the non-degenerate limit for the CB
and VB, each including an equivalent thermal generation. An example
is the SRH recombination model with a single trap level at energy
$\mathcal{E}_{I}$ \cite{Shockley52}, in which two trapping processes
(of an electron and a hole) produce a recombination event, with recombination
rate
\begin{align}
r^{\text{SRH}} & =\frac{pn-n_{i}^{2}}{(p+p_{1})\tau_{n}+(n+n_{1})\tau_{p}},\label{eq:SRH_r}
\end{align}
where $\tau_{p},\tau_{n}$ are the carrier lifetimes and $p_{1},n_{1}$
are the carrier concentrations of holes and electrons, respectively,
if their quasi-Fermi levels were equal to $\mathcal{E}_{I}$. This
$r^{\text{SRH}}$ appears as a negative contribution to $g_{k}$ for
both CB and VB.

We can model traps as intermediate bands with $w_{I}$ tracked explicitly,
in which case we implement standard Shockley-Read trapping \cite{Shockley52},

\begin{align}
r_{Ik}^{\text{SR}} & =\left[1-e^{s_{k}(w_{k}-w_{I})/k_{B}T}\right]f_{I,-s_{k}}u_{k}/\tau_{k},\label{eq:SR-trapping}
\end{align}
where $f_{I,-s_{k}}$ is the IB filling fraction of carriers with
charge $-s_{k}$, and $\tau_{k}$ is the Shockley-Read lifetime for
band $k$, as in Eq.~\ref{eq:SRH_r} \cite{Shockley52}. Note that
$r_{IC}^{\text{SR}}$ makes a negative contribution to $g_{C}$ and
a positive contribution to $g_{I}$, while $r_{IV}^{\text{SR }}$
makes a negative contribution to both $g_{V}$ and $g_{I}$.

Simudo also implements radiative trapping from band $k=C,V$ to $I$.
When we use Boltzmann statistics rather than Bose statistics for the
emitted photons, which is valid when $|w_{k}-w_{I}|$ remains at least
a few $k_{B}T$ below $|\mathcal{E}_{k}-\mathcal{E}_{I}|$, as in
Ref.~\cite{Strandberg11}, then the radiative trapping can be written
\begin{equation}
r_{Ik}^{\text{rad}}=\left[e^{-s_{k}(w_{k}-w_{I})/k_{B}T}-1\right]u_{I,s_{k}}\mathcal{I}_{Ik},\label{eq:rad-tmp}
\end{equation}
where 
\begin{align}
\mathcal{I}_{Ik} & =\frac{8\pi n_{r}^{2}}{h^{3}c^{2}}\,\int_{0}^{\infty}\sigma_{Ik}^{\text{opt}}(E)\,E^{2}e^{-E/k_{B}T}\,\d E,\label{eq:Strandberg-I-integral}
\end{align}
where $n_{r}$ is the index of refraction, and $u_{k,1}$ is either
$n_{1}$ or $p_{1}$ for $k=C,V$, respectively. Note that Ref.~\cite{Strandberg11}
includes only the recombination term, and we add the corresponding
thermal generation term, which is the -1 in Eq.~\ref{eq:rad-tmp}.
We can re-express Eq.~\ref{eq:rad-tmp} in a similar form to the
nonradiative terms by using the relation

\[
u_{I,s_{k}}=\frac{u_{I,-s_{k}}u_{k}}{u_{k,1}}e^{s_{k}(w_{k}-w_{I})/k_{B}T},
\]
which follows from Eqs.~\ref{eq:nondeg-u}, \ref{eq:deg-uI}. Then
\begin{equation}
r_{Ik}^{\text{rad}}=\left[1-e^{s_{k}(w_{k}-w_{I})/k_{B}T}\right]\,\frac{u_{I,-s_{k}}u_{k}}{u_{k,1}}\,\mathcal{I}_{Ik}.\label{eq:rad-trapping}
\end{equation}

As with Eq.~\ref{eq:nondeg-u}, Eqs.~\ref{eq:SR-trapping}-\ref{eq:Strandberg-I-integral}
are valid in the non-degenerate limit where $w_{k}$ does not approach
$\mathcal{E}_{k}$ but full degenerate statistics are used for the
IB. Extensions to the degenerate limit can be added, if desired. Simudo
also treats standard radiative recombination between conduction and
valence bands \cite{Nelson03}.

We also treat surface recombination at external surfaces $\Gamma$
of the device, which imposes a boundary condition 
\begin{equation}
\vec{j}_{k}\cdot\hat{n}\vert_{\Gamma}=S_{k}(u_{k}-u_{k0})\vert_{\Gamma},
\end{equation}
where $S_{k}$ is the surface recombination velocity of carriers in
band $k$ at boundary $\Gamma,$ $\hat{n}$ is the normal to $\Gamma$,
and $u_{k0}$ is the carrier concentration at equilibrium \cite{McIntosh14}.
The current release of Simudo supports only $S_{k}=0$ or $\infty$,
which impose $\vec{j}_{k}\cdot\hat{n}\vert_{\Gamma}=0$ or $(u_{k}-u_{k0})\vert_{\Gamma}=0$,
respectively.

\subsection{\label{subsec:opt}Optical equations }

For each wavelength, we need to solve the optical propagation according
to Eq.~\ref{eq:beerlambert}. For stability of the numerical solution,
it is convenient to use a second-order equation so that we can apply
boundary conditions on both the inlet and outlet boundaries \cite{Zhao2013a}.
We take the derivative of Eq.~\ref{eq:beerlambert} with respect
to the direction of propagation,
\begin{align}
\hat{s}\cdot\vec{\nabla}(\hat{s}\cdot\vec{\nabla}\Phi_{\lambda,\hat{s}})+\hat{s}\cdot\vec{\nabla}(\alpha_{\lambda}\Phi_{\lambda,\hat{s}}) & =0,\label{eq:opt-second-order}
\end{align}

With no reflection from the back, the boundary conditions are then
\begin{subequations}
\label{eq:opt-bcs}
\begin{alignat}{3}
\Phi_{\lambda,\hat{s}} & =\Phi_{\lambda,\hat{s}}^{0} & \qquad\vec{x} & \in\Gamma_{i} & \quad & \textnormal{(inlet)}\label{eq:opt-bc-d}\\
\hat{s}\cdot\vec{\nabla}\Phi_{\lambda,\hat{s}}+\alpha_{\lambda}\Phi_{\lambda,\hat{s}} & =0 & \vec{x} & \in\Gamma_{o} &  & \textnormal{(outlet)}\label{eq:opt-bc-n}
\end{alignat}
\end{subequations}
where $\Phi_{\lambda,\hat{s}}^{0}$ is a spectral photon flux at the
inlet boundary.

In the case where $\alpha_{\lambda}$ is constant for $\lambda$ in
an interval $[\lambda_{1},\lambda_{2}]$, the optical flux at all
wavelengths in that range obeys Eq.~\ref{eq:opt-second-order} and
can thus be treated together. We can write 
\[
\Phi_{[\lambda_{1},\lambda_{2}],\hat{s}}=\int_{\lambda_{1}}^{\lambda_{2}}\Phi_{\lambda,\hat{s}}\mathrm{d}\lambda,
\]
 where $\Phi_{[\lambda_{1},\lambda_{2}],\hat{s}}$ is a photon flux
(where $\Phi_{\lambda,\hat{s}}$ is a spectral photon flux). In this
case, we have
\[
\hat{s}\cdot\vec{\nabla}(\hat{s}\cdot\vec{\nabla}\Phi_{[\lambda_{1},\lambda_{2}],\hat{s}})+\hat{s}\cdot\vec{\nabla}(\alpha_{\lambda}\Phi_{[\lambda_{1},\lambda_{2}],\hat{s}})=0.
\]
Simudo uses this form, which allows simple treatment of piecewise
constant absorption coefficients with a small number of optical fields
$\Phi_{[\lambda_{1},\lambda_{2}],\hat{s}}$. When optical fields with
only one propagation direction $\hat{s}$ are considered, we write
the spectral flux density $\Phi_{\lambda}$ and the flux density $\Phi_{[\lambda_{1},\lambda_{2}]}$.

\subsection{Poisson's equation}

In electrostatics, Poisson's equation relates $\phi$, the charge
density $\rho$, and the permittivity $\varepsilon$,
\begin{align}
\vec{\nabla}\cdot(\varepsilon\,\vec{\nabla}\phi) & =-\rho.\label{eq:poisson}
\end{align}
It can also be split into two equations
\begin{subequations}
\label{eq:poisson-2eq}
\begin{align}
\vec{\nabla}\phi & =-\vec{E},\label{eq:poisson-2eq-gradphi}\\
\vec{\nabla}\cdot(\varepsilon\vec{E}) & =\rho,\label{eq:poisson-2eq-divE}
\end{align}
\end{subequations}
where $\vec{E}$ is the electric field.

The charge density $\rho$ is the sum of the static charge and the
mobile charge in each band. In an IB material, 
\begin{align}
\rho & =q[-n+p-N_{I}(f_{I}-f_{I,0})+N_{D}-N_{A}],\label{eq:charge}
\end{align}
where $N_{A}$, $N_{D}$ are the shallow acceptor and donor doping
concentrations, respectively, the mobile charge in the IB is $qN_{I}(f_{I}-f_{I,0})$,
with $f_{I,0}$ the IB filling fraction of the bulk IB material at
$T=0\u K$. For a donor-type IB $f_{I,0}=1$, and for an acceptor-type
IB $f_{I,0}=0$. Note that in writing the shallow dopant terms $N_{D}$
and $N_{A}$, we are assuming complete ionization of these impurities.

\section{\label{sec:numerical-method}Numerical method}

Simudo uses the finite element method (FEM) to solve the coupled Poisson/drift-diffusion
and optics problems, given by Eqs.~\ref{eq:dd-g},\ref{eq:j-mu-rel},\ref{eq:opt-second-order},
and \ref{eq:poisson-2eq}. The FEM method divides the simulation domain
$\Omega$ into cells, which are generally triangles in 2D, and enforces
a weak form of the desired PDE's with a set of test functions defined
on those cells, with boundary conditions applied on the domain boundary
$\Gamma$. The method is well-described in many reference texts \cite{Johnson1987a,Gockenbach2006a,Donea2003a}.
In this section, we detail the weak forms used for these coupled equations
and the solution method for the resulting nonlinear system. We benchmark
Simudo against the industry standard Synopsys Sentaurus commercial
simulator on a standard semiconductor problem to show the quality
of our results.

\subsection{\label{subsec:solution-method}Solution method}

The PDD problem is a coupled nonlinear system of PDEs, which we solve
iteratively using Newton's method as implemented in the FEniCS package.
The solution procedure is outlined in Fig.~\ref{fig:workflow}. The
goal is to find a solution $y=(\phi,\vec{E},w_{C},w_{I},w_{V},\vec{j}_{C},\vec{j}_{I},\vec{j}_{V})$
that satisfies Eqs.~\ref{eq:dd-g},\ref{eq:j-mu-rel},\ref{eq:poisson-2eq}
and associated $\Phi_{\lambda}$ that obeys Eq.~\ref{eq:opt-second-order}.
The optical problem is solved alongside the PDD problem in a self-consistent
manner. That is, the PDD subproblem produces the absorption coefficient
$\alpha(\lambda)$ (which, for processes involving the IB, depends
on the filling fraction). The optical subproblem is then solved using
these absorption coefficients, yielding a new photon flux $\Phi_{\lambda}$,
which is fed back into the PDD where it enters in the optical carrier
generation process, and the cycle iterates until a self-consistent
solution is found.

The convergence of Newton's method depends on the quality of the initial
guess. Steps 1 and 2 in Fig.~\ref{fig:workflow} are the pre-solver,
which is used once to make the initial guess for the main Newton solver,
illustrated in step 3. The full procedure is:

\begin{figure}
\noindent \begin{centering}
\includegraphics[width=1\columnwidth]{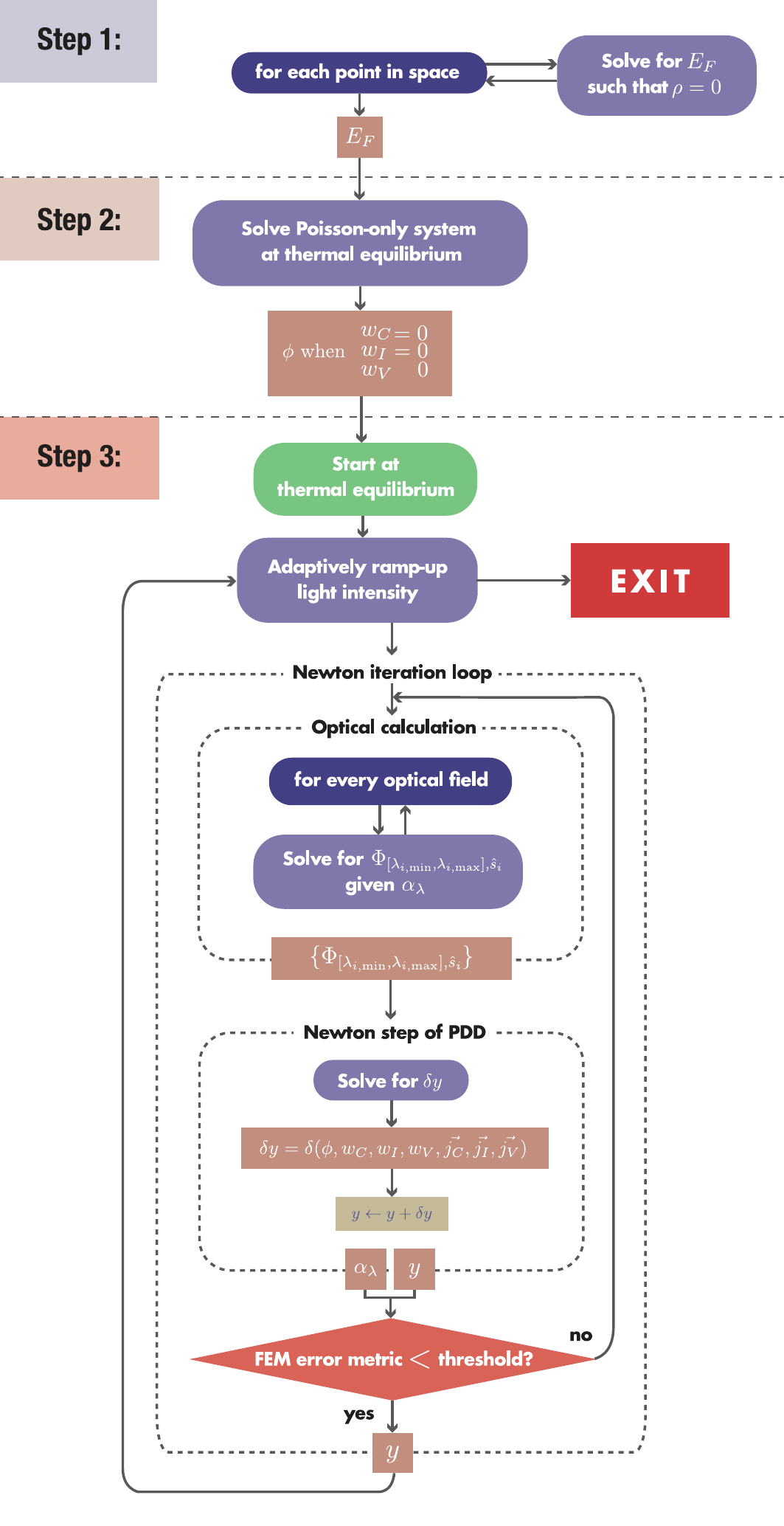}
\par\end{centering}
\caption{\label{fig:workflow}Workflow of the numerical method used in Simudo.
Steps 1 and 2 are pre-solver steps, which construct an initial guess
for the main Newton loop (step 3).}
\end{figure}

\begin{enumerate}
\item For each point in space, calculate the equilibrium Fermi level $E_{F}$
at that point assuming local charge neutrality and $\phi=0$.

Physically, this step finds the $E_{F}$ in each location that makes
it charge neutral, before any charge is allowed to flow.
\item Determine the built-in potential $\phi_{\mathrm{bi}}$ of the equilibrium
system, using $\phi_{\text{0}}=E_{F}/q$ as the initial guess. That
is, solve only Eq.~\ref{eq:poisson-2eq} for $\phi$ while keeping
all $w_{k}=0$ and thus all $\vec{j}_{k}=0$.

Physically, this step allows charge to move, forming depletion regions
as the carriers move to achieve a zero-current configuration that
satisfies Poisson's equation. The carrier density inside bulk-like
regions of space changes little from the bulk equilibrium value in
step 1, making $\phi_{\textnormal{0}}$ an excellent guess in large
regions of space.
\item \textbf{Main solver loop} Adaptively ramp up light intensity and/or
bias, starting with thermal equilibrium (dark, no bias). Each solution
requires a loop of Newton iterations. Within each Newton iteration,
do the following:
\begin{enumerate}
\item \textbf{Optical calculation }For each optical field $\lambda$, solve
for the photon flux $\Phi_{\lambda}$ given the latest value of $\alpha_{\lambda}$.
Note that Eq.~\ref{eq:opt-second-order} is linear when $\alpha_{\lambda}$
is fixed.
\item \textbf{PDD Newton step }Perform one Newton step of the PDD problem.
\begin{enumerate}
\item Solve for $\delta y=\delta(\phi,\vec{E},w_{C},w_{I},w_{V},\vec{j}_{C},\vec{j}_{I},\vec{j}_{V})$.
Use the value of $\Phi_{\lambda}$ (and thus optical carrier generation)
computed in the previous step.
\item Update $y\leftarrow y+\delta y$.

As an option, logarithmic damping can be applied to $\delta y$ to
prevent Newton's method from diverging, e.g., $y\leftarrow y+\textnormal{LogDamping}(\delta y)$
where $\textnormal{LogDamping}(z)=\k{sgn}(z)\log(1+c\abs z)/c$ for
$c=1.72$ or other user-defined value \cite{Gaury2019a}.
\end{enumerate}
\end{enumerate}
\end{enumerate}
We now describe the weak forms that we use for each of Eqs.~\ref{eq:dd-g},\ref{eq:j-mu-rel},
\ref{eq:opt-second-order}, and \ref{eq:poisson}. There is much flexibility
in the choice of particular weak forms, all of which can be equivalent
to the same strong form. In Section \ref{subsec:num-transport-qflop}
we illustrate the use of a partitioned offset representation for $w_{k}$,
which allows internal currents to be calculated accurately with double-precision
arithmetic.

\subsection{\label{subsec:num-poisson}Poisson equation}

Here we introduce the formulation we use to implement Eq.~\ref{eq:poisson-2eq}.
We use a mixed method to solve for both $\phi$ and $\vec{E}$ explicitly
\cite{Brezzi1985a,Roberts1991a}. The potential $\phi$ is represented
as a superposition of discontinuous Galerkin (DG) basis functions
of order $d_{\textnormal{poisson}}-1$ (cell-wise discontinuous polynomials),
 and $\vec{E}$ is represented using Brezzi-Douglas-Marini basis
functions of order $d_{\textnormal{poisson}}$ (cellwise discontinuous
polynomials with continuous normal component on cell boundaries) \cite{Brezzi1985a}.
We use the BDM space for all vector quantities, including $\vec{E}$
and $\vec{j}_{k}$. The BDM space is H(div) conforming, meaning the
divergence is accurately calculated and fluxes between cells are preserved,
which makes it a natural choice for conserved or almost conserved
vector quantities.\footnote{In the BDM space, the normal fluxes are shared by adjacent elements.
The flux exiting the perimeter of a collection of cells exactly equals
the sum of fluxes out of each of the cells, with exact arithmetic.} While $\vec{j}_{k}$ is not a conserved quantity, due to generation
and recombination that occur inside of cells, the BDM space ensures
that $\vec{j}_{k}$ is accurately preserved when passing between cells.
A method using CG or DG functions for $\vec{j}$ would be susceptible
to numerical errors associated with non-conservation of currents between
cells, and we show in Sec.~\ref{subsec:num-transport-qflop} that
Simudo conserves current well in a pn diode. In the results below,
$d_{\textnormal{poisson}}=2$.

We multiply Eq.~\ref{eq:poisson-2eq-gradphi} by test function $\vec{\psi}\in\mathrm{BDM}(d_{\textnormal{poisson}})$
and Eq.~\ref{eq:poisson-2eq-divE} by test function $v\in\mathrm{DG}(d_{\textnormal{poisson}}-1)$,
then integrate each spatially, giving  the weak forms
\begin{align}
\underbrace{\oint_{\Gamma}\vec{\psi}\cdot\hat{n}\,\phi_{\textnormal{BC}}}_{\textnormal{natural BC}}-\int_{\Omega}(\nabla\cdot\vec{\psi})\,\phi+\int_{\Omega}\vec{\psi}\cdot\vec{E} & =0\label{eq:num-poisson-gradphi}\\
\int_{\Omega}v\,\big(\nabla\cdot(\varepsilon\vec{E})\big)-\int_{\Omega}v\,\rho & =0,\label{eq:num-poisson-divE}
\end{align}
which must hold for every test function $\vec{\psi}$ and $v$, where
$\Omega$ is the full domain and $\Gamma$ is the boundary. Note that
Eq.~\ref{eq:num-poisson-gradphi} includes an integration by parts.
 In this case, the electric field BC is an essential BC, imposed
by reducing the set of test functions to those that satisfy the BC,
while the potential BC is a natural BC.

\subsection{\label{subsec:num-transport}Transport equations}

The drift-diffusion equations are often numerically challenging to
solve in semiconductors. In carrier density-based formulations, poor
resolution of the gradients of $u_{k}$ makes linear interpolation
of current density unstable, which the Scharfetter-Gummel box method
corrects for FVM methods \cite{Bank1983}. Additionally, catastrophic
cancellation can occur in Eq.~\ref{eq:dd-j}, e.g., for the majority
carrier in a quasi-neutral region of a semiconductor, when the drift
and diffusion contributions are nearly equal in magnitude. The current
is given by the difference and can be hard to evaluate with finite
precision arithmetic. We address these issues by using a quasi-Fermi-level-based
representation for carrier density \cite{Cummings09,Nachaoui99}.
Calculating $\vec{\nabla}w_{k}$ in finite precision for Eq.~\ref{eq:j-mu-rel}
can also be challenging when $w_{k}$ is very flat, and in Sec.~\ref{subsec:num-transport-qflop}
we introduce a partitioned offset representation for $w_{k}$ to allow
accurate determination of $\vec{\nabla}w_{k}$ with essentially no
extra computational cost. We use a mixed FEM method that solves explicitly
for both $w_{k}$ and the current density $\vec{j}_{k}$. As described
in Sec.~\ref{subsec:num-poisson}, the BDM space of basis functions
enforces local current conservation in the solutions, which also enables
local current densities to be well determined. Without the mixed method,
local current conservation is enforced indirectly, and we were not
able to obtain well-converged results for local currents.

\subsubsection{\label{subsec:num-transport-qfl}Quasi-Fermi level formulation}

The quasi-Fermi level $w_{k}$ is represented as a superposition of
DG basis functions of order $d_{\textnormal{transport}}-1$, and the
current density $\vec{j}_{k}$ is represented using BDM basis functions
of order $d_{\textnormal{transport}}$. Section~\ref{subsec:num-poisson}
contains a discussion of these functions' properties and of the mixed
method. In the results below, $d_{\textnormal{transport}}=2$.

We derive weak forms of Eq.~\ref{eq:dd-g} and Eq.~\ref{eq:j-mu-rel},
multiplying Eq.~\ref{eq:dd-g} by test function $v\in\mathrm{DG}(d_{\textnormal{transport}}-1)$,
taking the dot product of Eq.~\ref{eq:j-mu-rel} with the test function
$\vec{\psi}\in\mathrm{BDM}(d_{\textnormal{transport}})$, and integrating
each equation spatially, giving 
\begin{subequations}
\label{ddw}
\begin{align}
0 & =\int_{\Omega}v\,\vec{\nabla}\cdot\vec{j}_{k}-\int_{\Omega}s_{k}v\,qg_{k}\label{eq:ddw-g}\\
0 & =\int_{\Omega}\vec{\psi}\cdot\vec{j}_{k}/(\mu_{k}u_{k})-\underbrace{\oint_{\Gamma}(\vec{\psi}\cdot\hat{n})\,w_{k,\textnormal{BC}}}_{\textnormal{natural BC}}+\int_{\Omega}(\vec{\nabla}\cdot\vec{\psi})\,w_{k},\label{eq:ddw-j}
\end{align}
\end{subequations}
where the second equation was obtained by a further integration by
parts.

\subsubsection{\label{subsec:num-transport-qflop}Quasi-Fermi level offset partitioning}

As written, Eq.~\ref{eq:ddw-j} still suffers from a form of catastrophic
cancellation in its last term, which corresponds to the gradient term
in Eq.~\ref{eq:j-mu-rel}. Since $\int_{\Omega}\vec{\nabla}\cdot\vec{\psi}=0$
for $\vec{\psi}\in\text{BDM}(d_{\text{transport}})$, the last term
is nonzero only if $w_{k}$ varies within the domain where $\psi$
is nonzero. $w_{k}$ can be extremely flat, for example in quasineutral
regions, which makes this integral hard to calculate with finite arithmetic
precision. This difficulty is more apparent in Eq.~\ref{eq:j-mu-rel}:
if $\vec{j}_{k}/\mu_{k}u_{k}=\vec{\nabla}w_{k}$ is small,\footnote{relative to $|w_{k}|/(\textnormal{mesh size})$}
a representation of $w_{k}$ that stores its value on the nodes of
the mesh cannot resolve such small changes in $w_{k}$ across space.

We circumvent this issue by using an offset representation for $w_{k}$.
The idea is to give each cell in the domain its own (spatially constant)
base quasi-Fermi level $w_{k0}$ relative to which the new dynamical
variable $\delta w_{k}$ is expressed. That is, $w_{k}=w_{k0}+\delta w_{k}$
where $\delta w_{k}$ is the quantity we actually solve for instead
of $w_{k}$. Before every Newton iteration step, the $w_{k0}$ of
each cell is initialized to the cell average of $w_{k}$ from the
previous iteration. This representation allows small spatial changes
of $\delta w_{k}$ to be accurately represented, enabling accurate
determination of the current.

The last remaining question is how to adjoin regions with different
base $w_{k0}$ values. We connect them  by adding a surface integral
jump term to Eq.~\ref{eq:ddw-j}, resulting in
\begin{align}
0= & \int_{\Omega}\vec{\psi}\cdot\vec{j}_{k}/(\mu_{k}u_{k})-\oint_{\Gamma}(\vec{\psi}\cdot\hat{n})\,\delta w_{k,\textnormal{BC}}+\int_{\Omega}(\vec{\nabla}\cdot\vec{\psi})\,\delta w_{k}\nonumber \\
 & +\underbrace{\sum_{f\in\textnormal{interior facets}}\int_{f}(\vec{\psi}\cdot\hat{n})\,[w_{k0}]}_{\textnormal{region boundary term}}.\label{eq:ddw-j-delta}
\end{align}
where $[w_{k0}]$ is the jump operator, which takes the difference
between the values of a discontinuous expression on either side of
a facet. The rest of this section is dedicated to deriving that term
and comparing the result to a formulation without the offset representation.

We substitute $w_{k}=w_{k0}+\delta w_{k}$ into Eq.~\ref{eq:ddw-j},
and we obtain
\begin{align}
0= & \int_{\Omega}\vec{\psi}\cdot\vec{j}_{k}/(\mu_{k}u_{k})-\oint_{\Gamma}(\vec{\psi}\cdot\hat{n})\,\delta w_{k,\textnormal{BC}}+\int_{\Omega}(\vec{\nabla}\cdot\vec{\psi})\,\delta w_{k}\nonumber \\
 & \underbrace{-\oint_{\Gamma}(\vec{\psi}\cdot\hat{n})\,w_{k0,\textnormal{BC}}+\int_{\Omega}(\vec{\nabla}\cdot\vec{\psi})\,w_{k0}}_{(\star)}.\label{eq:ddw-j-delta-inter}
\end{align}

Our goal now is to rewrite the $(\star)$ term. Since $w_{k0}$ is
constant on each cell $K$, $\vec{\nabla}w_{k0}=\vec{0}$ within each
cell. We integrate by parts using

\begin{equation}
\int_{\Omega}\vec{\sigma}\cdot\vec{\nabla}a+\int_{\Omega}a(\vec{\nabla}\cdot\vec{\sigma})=\oint_{\Gamma}\vec{\sigma}a\cdot\hat{n},\label{eq:vector-integration-by-parts-1}
\end{equation}
yielding
\begin{align}
\int_{K}\vec{\psi}\cdot\underbrace{\vec{\nabla}w_{k0}}_{\vec{0}}+\int_{K}(\vec{\nabla}\cdot\vec{\psi})\,w_{k0} & =\oint_{\partial K}(\vec{\psi}\cdot\hat{n})\,w_{k0}
\end{align}

Summing over all cells $K$,
\begin{align*}
\underbrace{\sum_{K}\int_{K}}_{\int_{\Omega}}(\vec{\nabla}\cdot\vec{\psi})\,w_{k0}= & \sum_{K}\oint_{\partial K}(\vec{\psi}\cdot\hat{n})\,w_{k0}\\
\int_{\Omega}(\vec{\nabla}\cdot\vec{\psi})\,w_{k0}=\oint_{\Gamma}(\vec{\psi}\cdot\hat{n})\,w_{k0}+ & \sum_{f\in\textnormal{interior facets}}\int_{f}(\vec{\psi}\cdot\hat{n})\,[w_{k0}]\\
\underbrace{-\oint_{\Gamma}(\vec{\psi}\cdot\hat{n})\,w_{k0}+\int_{\Omega}(\vec{\nabla}\cdot\vec{\psi})\,w_{k0}}_{(*)}= & \sum_{f\in\textnormal{interior facets}}\int_{f}(\vec{\psi}\cdot\hat{n})\,[w_{k0}]
\end{align*}
which, plugged into Eq.~\ref{eq:ddw-j-delta-inter}, yields Eq.~\ref{eq:ddw-j-delta}.

\begin{figure}
\noindent \begin{centering}
\includegraphics[width=1\columnwidth]{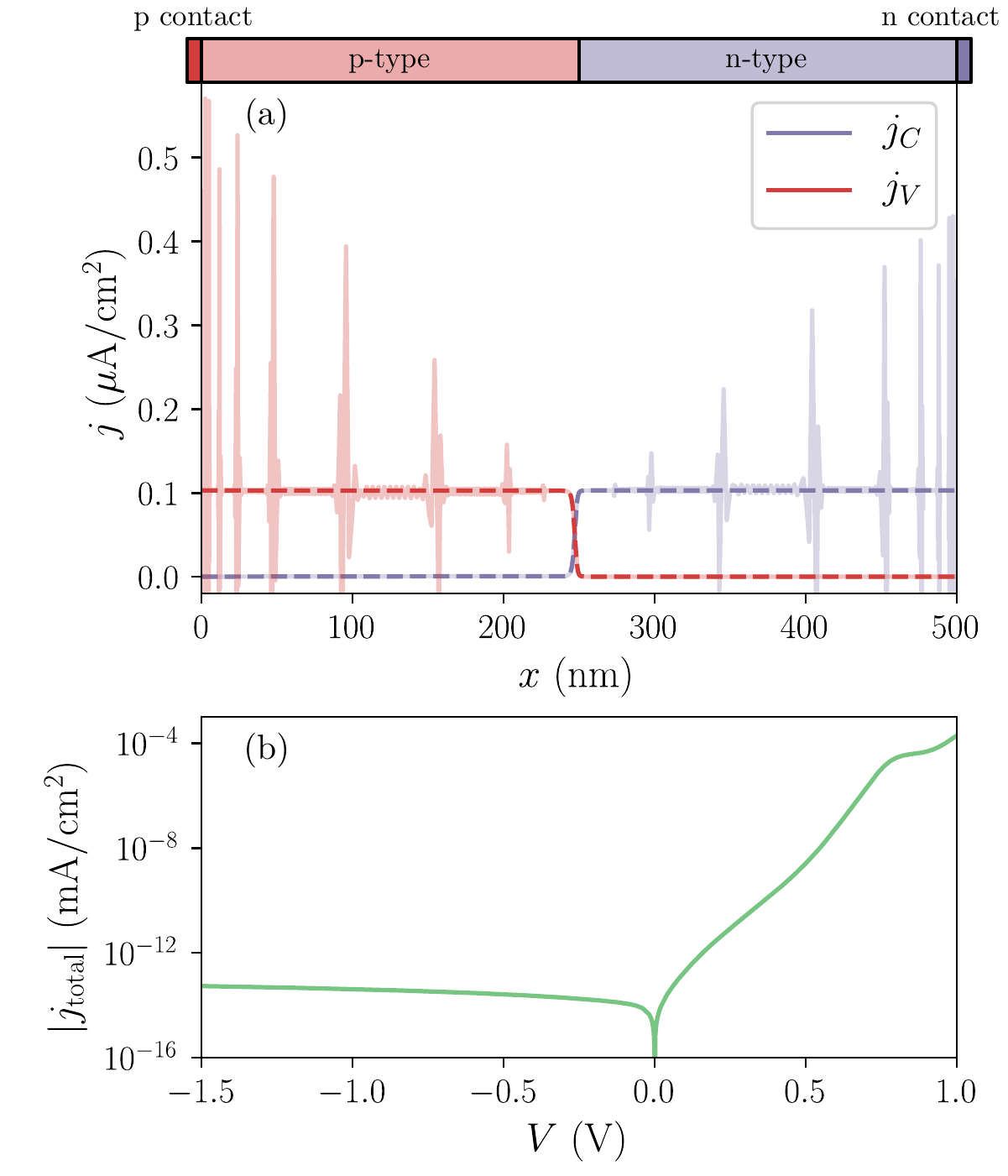}
\par\end{centering}
\caption{\label{fig:why-constant-offset} (a) Current densities $j_{C}$ and
$j_{V}$ for a silicon pn-junction at a bias of 0.16~V. Simudo's
partitioned offset representation (dashed, dark lines) allows robust
determination of internal current densities, while calculations without
the offset representation (solid, dull lines) cannot determine the
majority current densities accurately. (b) Total charge current density
at the contact. Results with and without the offset representation
agree to 5 digits of precision.}
\end{figure}

We perform a test of Simudo, which uses the partitioned offset representation,
against the identical model without the offset representation. We
consider a standard silicon pn-junction diode with symmetric doping
of $10^{18}$~cm$^{-3}$ and SRH lifetimes of 1~ns and 1~$\mu$s
in the p- and n-type regions, respectively. Each region has a length
of 250~nm, for total device length of 500~nm. Although the problem
is one-dimensional, we consider a 2D region with a height of $1\u{\mu m}$.
At each contact, the majority carrier has an infinite surface recombination
velocity while the minority carrier has zero surface recombination.
We use a mesh with 769 points in the x-direction, which is tightest
near the contacts and junction and expands out geometrically toward
the middle of the quasi-neutral regions. The mesh has 2 points in
y-direction, and further details of the mesh are given in Sec.~\ref{subsec:num-transport-sentaurus}.

The offset representation allows internal current densities to be
resolved accurately throughout the device. Figure \ref{fig:why-constant-offset}(a)
shows the electron and hole currents under 0.16~V bias. Without the
offset representation, the majority currents are poorly resolved,
due to the inability to resolve $\vec{\nabla}w_{k}$ with double-precision
arithmetic. The majority currents in the no-offset model become worse
as the mesh density increases (not shown), as expected for approximations
of $\vec{\nabla}w_{k}$. Figure \ref{fig:why-constant-offset}(b)
shows the total charge current density at the contact, and the results
with and without the offset representation agree to 5 digits. We conclude
that the offset representation allows robust extraction of internal
current densities but does not seem to be important for the overall
current density of the test device. The offset representation imposes
essentially no extra computational cost on Simudo while enabling robust
determination of internal current densities.

\subsection{Optics}

The optical problem is solved by self-consistently iterating through
the optical flux variables $\Phi_{[\lambda_{i,\textnormal{min}},\lambda_{i,\textnormal{max}}],\hat{s}_{i}}$
and independently solving Eq.~\ref{eq:opt-second-order} for each
one. For convenience, we write $\Phi=\Phi_{[\lambda_{i,\textnormal{min}},\lambda_{i,\textnormal{max}}],\hat{s}_{i}}$,
$\hat{s}=\hat{s}_{i}$, and $\alpha=\alpha_{[\lambda_{i,\textnormal{min}},\lambda_{i,\textnormal{max}}]}$
for the remainder of this section. We represent $\Phi$ using CG basis
functions of order $d_{\textnormal{optical}}=2$.

We now derive the weak form used in Simudo to solve each optical propagation
problem. We follow closely the derivation in \cite{Zhao2013a} of
the modified second order radiative transfer equation (MSORTE) method,
without the scattering matrix. Integrating Eq.~\ref{eq:opt-second-order}
with a test function $v\in\mathrm{CG}(d_{\textnormal{optical}})$
gives
\begin{align}
\int_{\Omega}\vec{v}\cdot\vec{\nabla}\zeta+\int_{\Omega}\vec{v}\cdot\vec{\nabla}(\alpha\,\Phi) & =0,
\end{align}
where $\vec{v}=v\,\hat{s}$ and $\zeta=\hat{s}\cdot\vec{\nabla}\Phi$.
Using Eq.~\ref{eq:vector-integration-by-parts-1}, we obtain 
\begin{align}
\oint_{\Gamma}(\vec{v}\cdot\hat{n})\,\zeta_{\textnormal{BC}}-\int_{\Omega}(\vec{\nabla}\cdot\vec{v})\,\zeta+\int_{\Omega}\vec{v}\cdot\vec{\nabla}(\alpha\,\Phi) & =0.
\end{align}

Inserting the outlet boundary condition Eq.~\ref{eq:opt-bc-n} into
the first term, we obtain the final weak form
\begin{align}
\oint_{\Gamma}(\vec{v}\cdot\hat{n})\,(-\alpha\,\Phi)-\int_{\Omega}(\vec{\nabla}\cdot\vec{v})\,\zeta+\int_{\Omega}\vec{v}\cdot\vec{\nabla}(\alpha\,\Phi) & =0
\end{align}

The inlet boundary condition Eq.~\ref{eq:opt-bc-d} is applied directly
on $\Phi$ as an essential boundary condition.

\subsection{\label{subsec:num-transport-sentaurus}Sentaurus benchmark comparison}

To validate Simudo, we benchmark it against the industry standard
Synopsys Sentaurus device simulator. Since Sentaurus does not support
intermediate band materials, the benchmark is limited to standard
semiconductors. Our test problem is the same silicon pn-junction as
considered in Section \ref{subsec:num-transport-qflop}, with the
overall $J(V)$ as shown in Fig.~\ref{fig:why-constant-offset}(b).
See below for a discussion of the differences in implementation of
the Ohmic condition between Simudo and Sentaurus. 

We study the convergence of the results as the mesh is refined, using
a number of mesh points in the x-direction ranging from 45 to 12289,
with the same meshes used for both Sentaurus and Simudo. The mesh
spacing is nonuniform in the x-direction since the carrier and current
densities vary most rapidly near the contacts and the junction. The
meshes are generated by splitting the structure into two regions,
one extending from the p-contact to the junction and one from the
n-contact to the junction. In each region, a mesh spacing $d_{0}$
is applied to the cell adjacent to the contact and the cell adjacent
to the junction. The mesh spacing increases geometrically toward the
center of each region with a growth factor of 1.2. To generate the
finest mesh sizes, these cells are further subdivided into 4, 16 or
64 equal parts. There are 2 points in the y-direction. The computational
cost generally increases with the number of degrees of freedom rather
than with the number of cells in the mesh. Simudo has more degrees
of freedom associated with each cell than Sentaurus, due to its higher-order
basis functions. For this mesh, Simudo has 36 degrees of freedom per
triangle while Sentaurus has 9 per triangle, with 2 triangles per
mesh point. In this mesh, each triangle has at least one edge on the
boundary of the device, which increases the number of degrees of freedom
per triangle compared to a mesh where most triangles share sides;
this effect is similar for both methods.

\begin{figure}
\includegraphics[width=1\columnwidth]{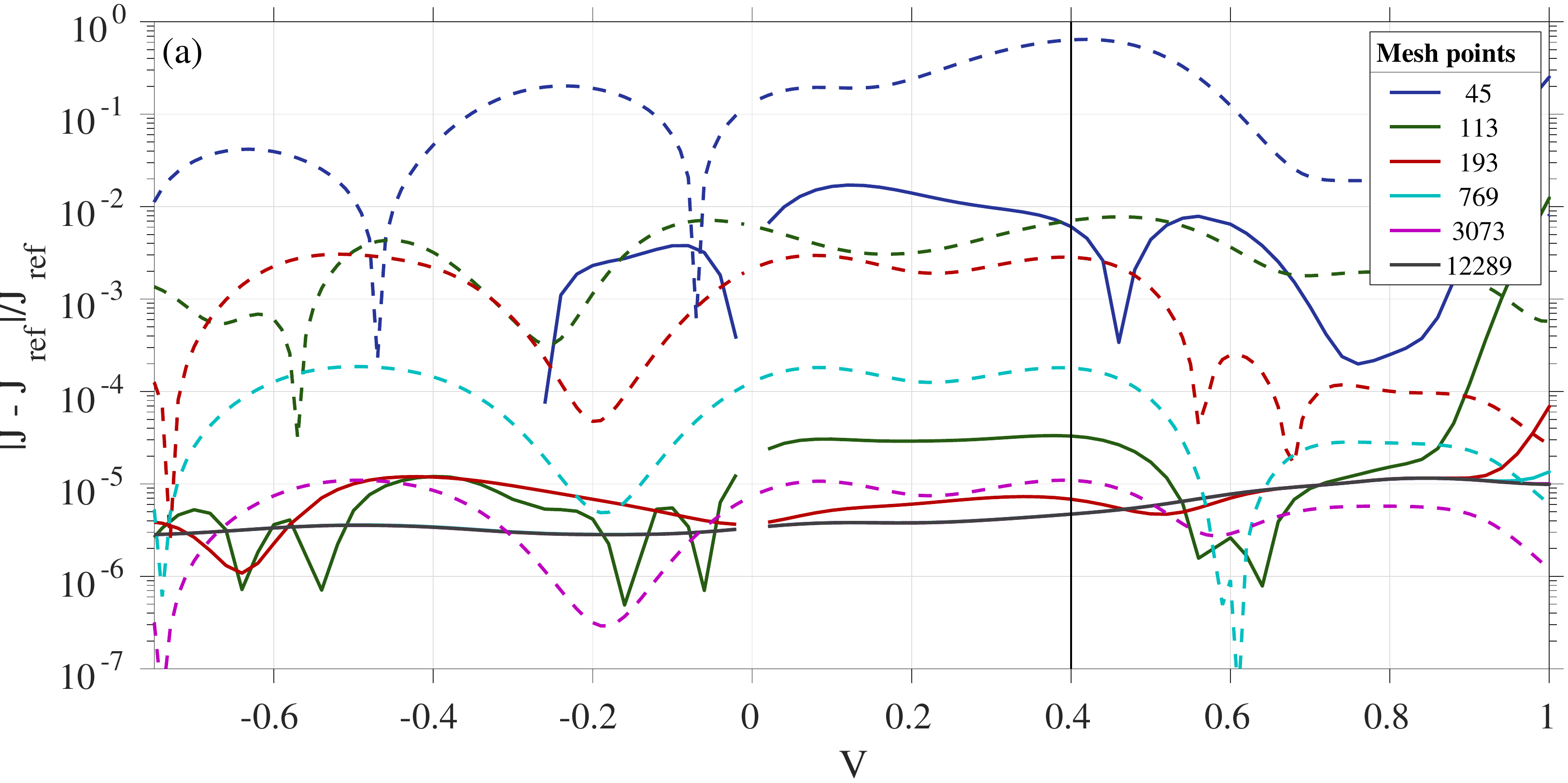}

\includegraphics[width=1\columnwidth]{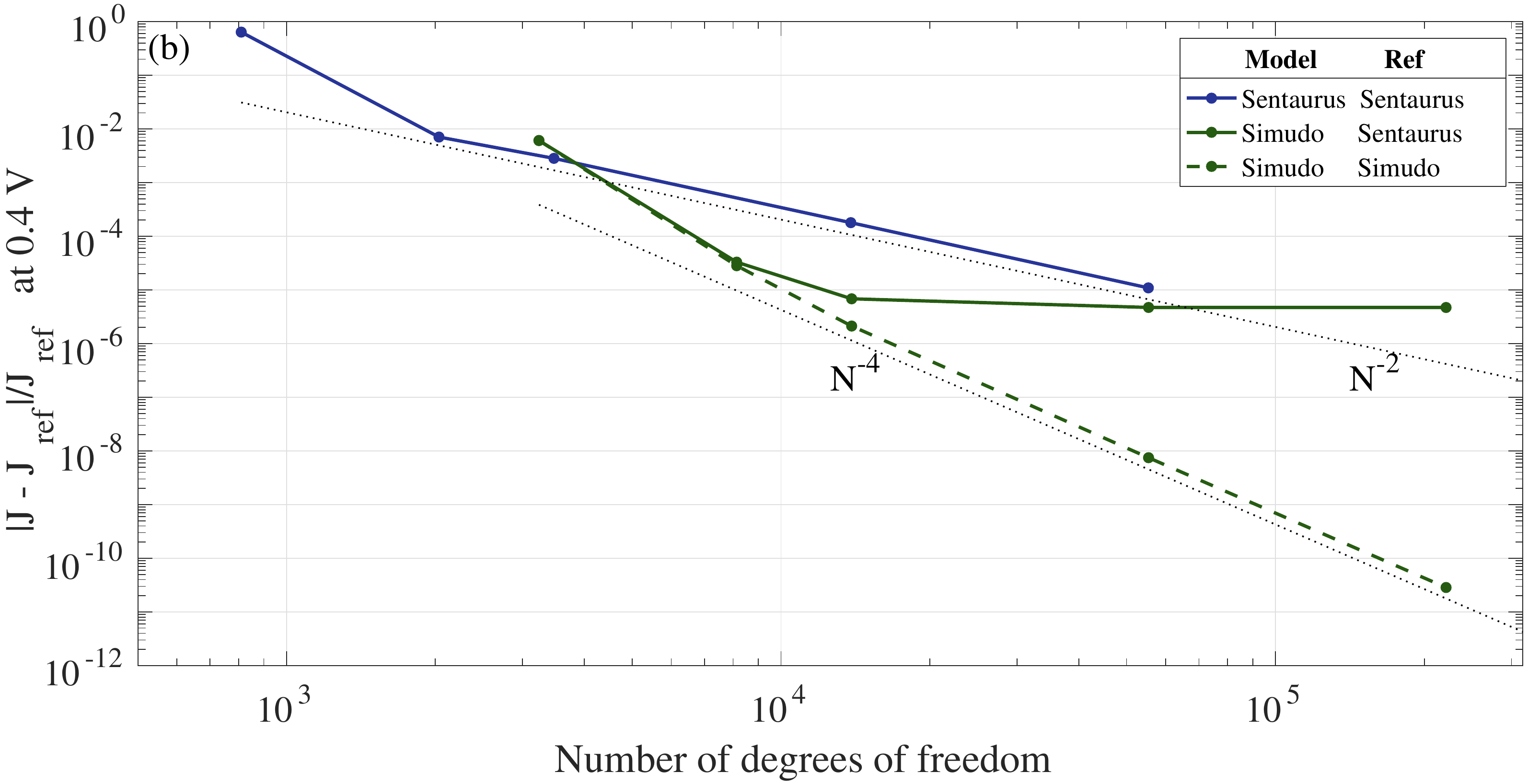}

\caption{(a) Relative error in total current $J$ for a test pn-junction in
the dark, at several mesh densities. Dashed lines show results from
Sentaurus, and solid lines show results from Simudo, with the same
meshes. The Sentaurus result with the finest mesh is taken as $J_{\text{ref}}$
for calculating relative error. Points at $V=0$ are removed to avoid
dividing by zero. (b) Relative errors at 0.4~V plotted against the
number of degrees of freedom in each simulation. Solid lines use Sentaurus
results for $J_{\text{ref}}$, and the dashed line shows Simudo self-convergence.
Dotted lines show the scaling trends of the two methods.\label{fig:Sentaurus-mesh}}
\end{figure}

Figure \ref{fig:Sentaurus-mesh}(a) shows the results of the study,
where we plot the relative error in $J(V)$ for each simulation, with
the reference current $J_{\text{ref}}$ taken from Sentaturus with
the densest mesh. With 193 mesh points, Simudo converges approximately
as well as Sentaurus with 3073 mesh points. Above 769 mesh points,
the Simudo results show no further improvement in error, indicating
either that Simudo and Sentaurus converge to results that differ at
the $10^{-5}$ level or that Sentaurus with 12289 mesh points is only
converged to the $10^{-5}$ level, while Simudo may have converged
more precisely. The simulations were performed on different machines,
so we do not report timing data.

The figure makes clear that Simudo converges much more rapidly with
mesh size than Sentaurus, which demonstrates the higher order convergence
that FEM is supposed to provide over FVM. Figure \ref{fig:Sentaurus-mesh}(b)
shows the scaling of the errors with the number of degrees of freedom
in each simulation, at 0.4~V. The solid blue line shows that Sentaurus'
self-convergence scales like $N^{-2}$ with the number of degrees
of freedom. The dashed green line shows Simudo's self-convergence,
with $J_{\text{ref}}$ taken from the Simudo simulation with 12289
mesh points. It shows that Simudo's self-convergence scales like $N^{-4}$
with the number of degrees of freedom. For all but the smallest meshes,
Simudo's convergence is superior to Sentaurus' at the same number
of degrees of freedom. Taken together, these figures show that with
193 mesh points, Simudo's result is as good as Sentaurus' with 64
times as many mesh points, which is equivalent to 16 times as many
degrees of freedom.

Note that the boundary conditions at the contacts are not precisely
the same for the Sentaurus and Simudo simulations. Both are intended
to simulate Ohmic contacts for the majority carrier and surface recombination
velocities of 0 for the minority carrier. The Simudo simulations are
performed with surface recombination velocity $S=\infty$ and 0 for
the majority and minority carriers, respectively, imposing equilibrium
carrier concentration at the boundary for majority carriers and setting
$\vec{j}_{k}\cdot\hat{n}\vert_{\Gamma}=0$ for minority carriers,
as described in Sec.~\ref{subsec:Recombination-and-trapping}. The
Sentaurus simulations are performed with $S=0$ for the minority carriers,
in agreement with Simudo, and the default ``Ohmic contact'' boundary
condition for the majority carriers, which imposes charge neutrality
and equilibrium carrier concentration at the contact. Under small
and reverse bias, these two sets of boundary conditions should be
equivalent, but under large forward bias, the default Sentaurus boundary
condition is expected to give incorrect results due to its imposition
of charge neutrality \cite{Synopsys15}. Sentaurus provides a ``Modified
Ohmic'' boundary condition, which should be closer to the Simudo
boundary condition, but we were unable to attain convergence using
it. As a result, at larger biases the Simudo and Sentaurus results
diverge from each other, and we do not include them in Fig.~\ref{fig:Sentaurus-mesh}.
For biases larger than 1~V, the diode is in high injection, and the
Boltzmann approximation used in this calculation is not accurate,
regardless.

\section{Examples and results\label{sec:Examples-and-results}}

In this section, we give examples of using Simudo. Section \ref{subsec:PN-junction-example}
shows how to set up a simple 1-dimensional pn-junction device and
demonstrates the helpful tools that Simudo provides for defining regions
and boundaries. Section \ref{subsec:Auger_recombination} shows the
extensibility of Simudo by illustrating the code required to add a
new Auger recombination process. Section \ref{subsec:Marti_comparison}
illustrates the use of Simudo to study a system first considered in
Ref.~\citenum{Marti02}.

\subsection{PN junction and topology definitions\label{subsec:PN-junction-example}}

We include in the supplementary material the code listing $\texttt{equilibrium.py}$
describing a simple pn-junction device in Simudo. This example constructs
the device and implements steps 1-4 of the pre-solver shown in Fig.~\ref{fig:workflow}.
Here, we discuss some of the pieces of that code and illustrate the
useful topology construction operations built in to Simudo.

\begin{figure}
\begin{centering}
\includegraphics[width=9cm]{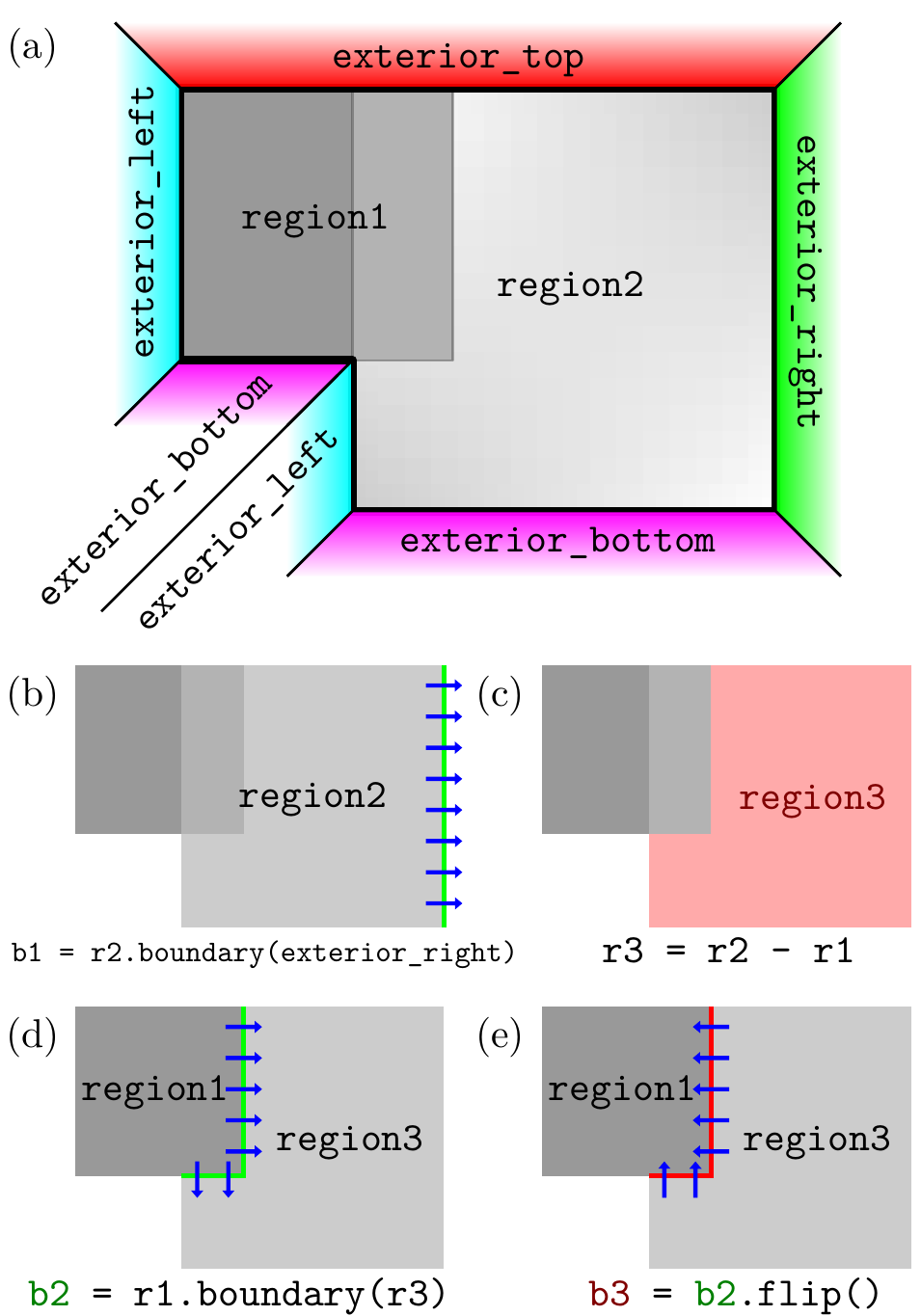}
\par\end{centering}
\caption{\label{fig:topology-drawing} Simudo allows users to define complicated
domains and regions within them, with useful tools to automatically
keep track of the required cells and facets in each region. (a) A
simulation domain divided into overlapping $\texttt{region1}$ and
$\texttt{region2}$. Simudo automatically defines the $\texttt{exterior}$
outside the simulation domain, and provides the helper regions $\texttt{exterior\_left}$,
$\texttt{exterior\_right}$, $\texttt{exterior\_top}$, and $\texttt{exterior\_bottom}$,
which may not be helpful, depending on the shape of the domain. Even
before the regions are given specific geometries in the simulation
domain (e.g., those pictured in (a)), Simudo's topology tools allow
construction of further regions and facets. (b-e) Derived regions
and facets from the domain shown in (a), with the Simudo command to
produce them shown underneath, where we shorten $\texttt{regionX}$
to $\texttt{rX}$. Further example code is shown in Fig.~\ref{fig:topology-example-code}.
(b) The contact $\texttt{FacetRegion}$ $\texttt{b1}$ is the signed
boundary from $\texttt{region2}$ to $\texttt{exterior\_right}$,
with the sign indicated by the arrows. It can be used for determining
the current flow out of the device. (c) $\texttt{region3}$ is $\texttt{region2}$
with $\texttt{region1}$ removed. (d) The $\texttt{FacetRegion}$
$\texttt{b2}$ is the signed boundary from $\texttt{region1}$ to
$\text{\texttt{region3}}$. (e) The $\texttt{flip}$ operation reverses
the sign of the boundary.}
\end{figure}
\begin{figure}
\begin{lstlisting}[language=Python]
from simudo.mesh import CellRegions, FacetRegions
R = CellRegions() ; F = FacetRegions()

# R.region1 is created automatically when referenced
print(R)
print(R.region1, type(R.region1))
print(R)
## CellRegions({})
## CellRegionByName('region1') <class 'simudo.mesh.topology.CellRegionByName'>
## CellRegions({'region1': CellRegionByName('region1')})

# Define new region as set difference of region2 and region1
R.region3 = R.region2 - R.region1
# The method (R.region1).boundary(R.region2) creates the signed boundary from region1 into region2, e.g., for calculating flux. R.region2 is created when referenced.
F.b1 = (R.region2).boundary(R.exterior_right)
F.b2 = (R.region1).boundary(R.region3)
F.b3 = (F.b2).flip()
\end{lstlisting}

\caption{\label{fig:topology-example-code}Example of using $\texttt{CellRegions}$
and $\texttt{FacetRegions}$ to construct regions corresponding to
those shown in Fig.~\ref{fig:topology-drawing}. The $\texttt{R}$
and $\texttt{F}$ objects are containers for cell and facet regions
respectively. They are both initialized empty, and regions in each
one are created when referenced. The $\texttt{boundary}$ method in
$\texttt{(R.region1).boundary(R.region3)}$ creates the signed boundary
from $\texttt{region1}$ to $\texttt{region3}$. The $\texttt{flip()}$
method gives the boundary with the opposite sign. The actual mapping
of these regions into the domain occurs in the mesh generation.}
\end{figure}

In the 1-dimensional pn-junction example, the object $\texttt{ls}$
contains information about the layers, including their sizes, positions,
and mesh. The object $\texttt{pdd}$ sets up the Poisson/drift-diffusion
solver and has information about the bands in each material, including
recombination processes and boundary conditions. In this example,
there are only two bands (VB, CB); for a problem including an IB,
$\texttt{pdd}$ would have a third band, too.

Simudo is designed for 2-dimensional simulations, and it has sophisticated
tools to define the arrangement of materials, dopings, contacts, meshing
regions, or other user-defined spatial properties. In many FEM solvers,
interfaces must be tracked manually, including their orientation,
to ensure that integrals over those interfaces are added together
properly. Simudo introduces a set of topology tools that instead allow
users to define the regions and interfaces in which they are interested,
and Simudo takes care of all the bookkeeping. The user defines regions
as desired (e.g., emitter, base, defective-region), which can then
be given properties, whether they be doping levels, recombination
parameters, or other desired properties. These regions are initially
defined abstractly, without having any coordinates in the device,
using \texttt{$\texttt{CellRegions}$} and \texttt{$\texttt{FacetRegions}$},
and are later connected to geometry and materials by the mesh generator.

Full details are given in the documentation accompanying Simudo, but
we give a further illustration of these methods in Figs.~\ref{fig:topology-drawing}
and \ref{fig:topology-example-code}. That example illustrates the
creation of arbitrary $\texttt{CellRegion}$ objects, including unions
and intersections, and edges that connect them. When $\texttt{R}$
is a $\texttt{CellRegions}$ container, accessing a nonexistent attribute
(such as $\texttt{R.domain}$) causes its creation. The user can define
new $\texttt{CellRegion}$ objects by applying Boolean operations
on previous ones and new $\texttt{FacetRegion}$ objects by using
the $\texttt{boundary}$ method. For example, consider the region
$\texttt{R.region1}$. Then \\
$\texttt{R.region1.boundary(R.region2)}$ creates a signed boundary
from $\texttt{region1}$ to $\texttt{region2}$, as illustrated in
Fig.~\ref{fig:topology-drawing}. All of these custom regions are
kept as symbolic expressions and evaluated by Simudo only when needed
(e.g., when asked to apply a boundary condition or when asked to compute
a volume or surface integral). This layer of abstraction allows the
user not to worry about the details of mesh markers, entity indices,
and facet orientations \cite{LoggMardalEtAl2012a}, and is described
more fully in the documentation that accompanies Simudo.

The examples in Figs.~\ref{fig:topology-drawing}-\ref{fig:topology-example-code}
illustrate another useful concept. The mesh generation interprets
the $\texttt{external}$ region as being outside the simulation domain,
allowing convenient definitions for boundary conditions and current
flow. The $\texttt{FacetRegions}$ are used in the pn-junction example
shown in the supplementary material to define the boundary conditions,
which -- in step 2 -- are conductive at the left and right contacts
and nonconductive at the top and bottom surfaces. That example also
shows how the mesh can be refined by adding extra mesh points near
the contacts.

\subsection{Extensibility: Adding Auger recombination\label{subsec:Auger_recombination}}

\begin{figure}

\begin{lstlisting}[language=Python]
from simudo.physics import (DarkEOPMixin, TwoBandEOPMixin, ElectroOpticalProcess)

class AugerRecombination(DarkEOPMixin, TwoBandEOPMixin, ElectroOpticalProcess):
  """Auger recombination"""
  name = 'Auger'

  def get_Auger_C(self, band):
    return self.pdd.spatial.get('/'.join((
           self.name, band.name, 'C')))

  def get_generation_user(self, band):
    # In this case, sign is +1 for CB, VB and None for all other bands
    sign = self.get_band_generation_sign(band)
    if sign is None:
      return self._zero_generation

    CB = self.dst_band ; C_n = self.get_Auger_C(CB)
    VB = self.src_band ; C_p = self.get_Auger_C(VB)

    n = CB.u ; n0 = CB.thermal_equilibrium_u
    p = VB.u ; p0 = VB.thermal_equilibrium_u

    r = (C_n*(n**2 * p - n0**2 * p0) +
         C_p*(p**2 * n - p0**2 * n0))

    return (-r) * sign  # Contribution to generation rate

...

spatial.add_rule('Auger/CB/C', R.domain,
     U('1.1e-30 cm^6/s'))
spatial.add_rule('Auger/VB/C', R.domain,
     U('0.3e-30 cm^6/s'))
\end{lstlisting}

\caption{\label{fig:auger_example_code} Example Simudo code that implements
Auger recombination (Eq.~\ref{eq:auger}), including the $\texttt{AugerRecombination}$
class. The last four lines show the Auger material parameters being
set throughout the domain.}
\end{figure}

The initial release of Simudo contains radiative and Shockley-Read
trapping and recombination processes in the non-degenerate limits
for VB, CB. The user can easily add modified physics to their problems,
which we demonstrate here with an example of adding an Auger recombination
process to Simudo, with the form 
\begin{align}
U_{A} & =C_{n}(n^{2}p-n_{0}^{2}p_{0})+C_{p}(p^{2}n-p_{0}^{2}n_{0}),\label{eq:auger}
\end{align}
where $C_{n},C_{p}$ are the Auger coefficients, and $p_{0}$ and
$n_{0}$ are the hole and electron concentrations at thermal equilibrium,
respectively \cite{Nelson03}. The code is listed in Fig.~\ref{fig:auger_example_code}.
The function $\texttt{get\_generation\_user(band)}$ adds a negative
local generation rate in the CB and VB and returns 0 for all other
bands. This recombination process moves particles between two bands,
the $\texttt{src\_band}$ and the $\texttt{dst\_band}$. In this case,
where the electrons and holes have opposite charge, the Auger process
destroys both particles simultaneously; if both carrier types involved
in the process had the same charge (e.g., for a CB-to-IB trapping
process), the process would represent a particle-number-conserving
transfer (rather than a recombination) from the $\texttt{src\_band}$
to the $\texttt{dst\_band}$, with the appropriate sign for the recombination
process determined by the $\texttt{get\_band\_generation\_sign}$
method (inherited from $\texttt{TwoBandEOPMixin}$). This method's
sign convention is that the $\texttt{dst\_band}$ always gains carriers
through the generation process, while the $\texttt{src\_band}$ gains
or loses as required by conservation of charge.

\subsection{P{[}IB{]}N junction\label{subsec:Marti_comparison}}

In Ref.~\cite{Marti2002a}, the authors consider a quantum-dot-based
IB solar cell with a p-n-IB-p-n structure. They present a drift-diffusion
model for the IB region only, with the carrier density and current
density boundary conditions obtained from a depletion approximation
and law of the junction. This model assumes that transport is diffusion-dominated
in the IB region, and drift can therefore be neglected. This early
device model gave important insights into the behavior of IB devices.

\begin{table}[h]
\caption{\label{tab:marti-param}Parameters modeled on the device from \cite{Marti2002a},
for Figs.~\ref{fig:marti-JV-mis}--\ref{fig:marti-f-IB}.}
\begin{tabular}{ll}
\hline 
\textbf{Value} & \textbf{Definition}\tabularnewline
\hline 
$\mathcal{E}_{C}=1.67\u{eV}$ & Conduction band edge energy\tabularnewline
$\mathcal{E}_{I}=1.10\u{eV}$ & Intermediate band energy\tabularnewline
$\mathcal{E}_{V}=0\u{eV}$ & Valence band edge energy\tabularnewline
$N_{C}=N_{V}=5\times10^{18}\u{cm^{-3}}$ & CB and VB effective density of states\tabularnewline
$N_{I}=10^{17}\u{cm^{-3}}$ & IB density of states\tabularnewline
$\mu_{C}=\mu_{V}=2000\u{cm^{2}/V/s}$ & CB and VB mobility\tabularnewline
$\mu_{I}=0.001-300\u{cm^{2}/V/s}$ & IB mobility\tabularnewline
$\alpha_{CV}=10^{4}\u{cm^{-1}}$ & Absorption coefficient for CV process\tabularnewline
$\sigma_{CI}^{\textnormal{opt}}=(2-10)\times10^{-13}\u{cm^{2}}$ & Optical cross section for CI process\tabularnewline
$\sigma_{IV}^{\textnormal{opt}}=2\times10^{-13}\u{cm^{2}}$ & Optical cross section for IV process\tabularnewline
$\varepsilon=13\varepsilon_{0}$ & Dielectric constant\tabularnewline
$T_{s}=6000\u K$ & Sun temperature\tabularnewline
$T_{c}=300\u K$ & Cell temperature\tabularnewline
$X=1000$ & Solar concentration factor\tabularnewline
$w_{\textnormal{IB}}=1.3\:\mu\mathrm{m}$ & IB region length\tabularnewline
$f_{I,0}=1/2$ & Charge-neutral IB filling fraction\tabularnewline
\end{tabular}
\end{table}

In testing the self-consistency of the model, the authors estimate
the IB mobility required to remain in the diffusion-dominated regime,
finding that an IB mobility greater than $62\u{cm^{2}/V/s}$ is required
to make their model consistent. This claim raises an immediate question:
does something interesting happen when the IB mobility goes below
that threshold? Since Simudo is a full drift-diffusion device model,
we can directly answer that question.

\begin{figure}
\noindent \begin{centering}
\includegraphics[width=9cm]{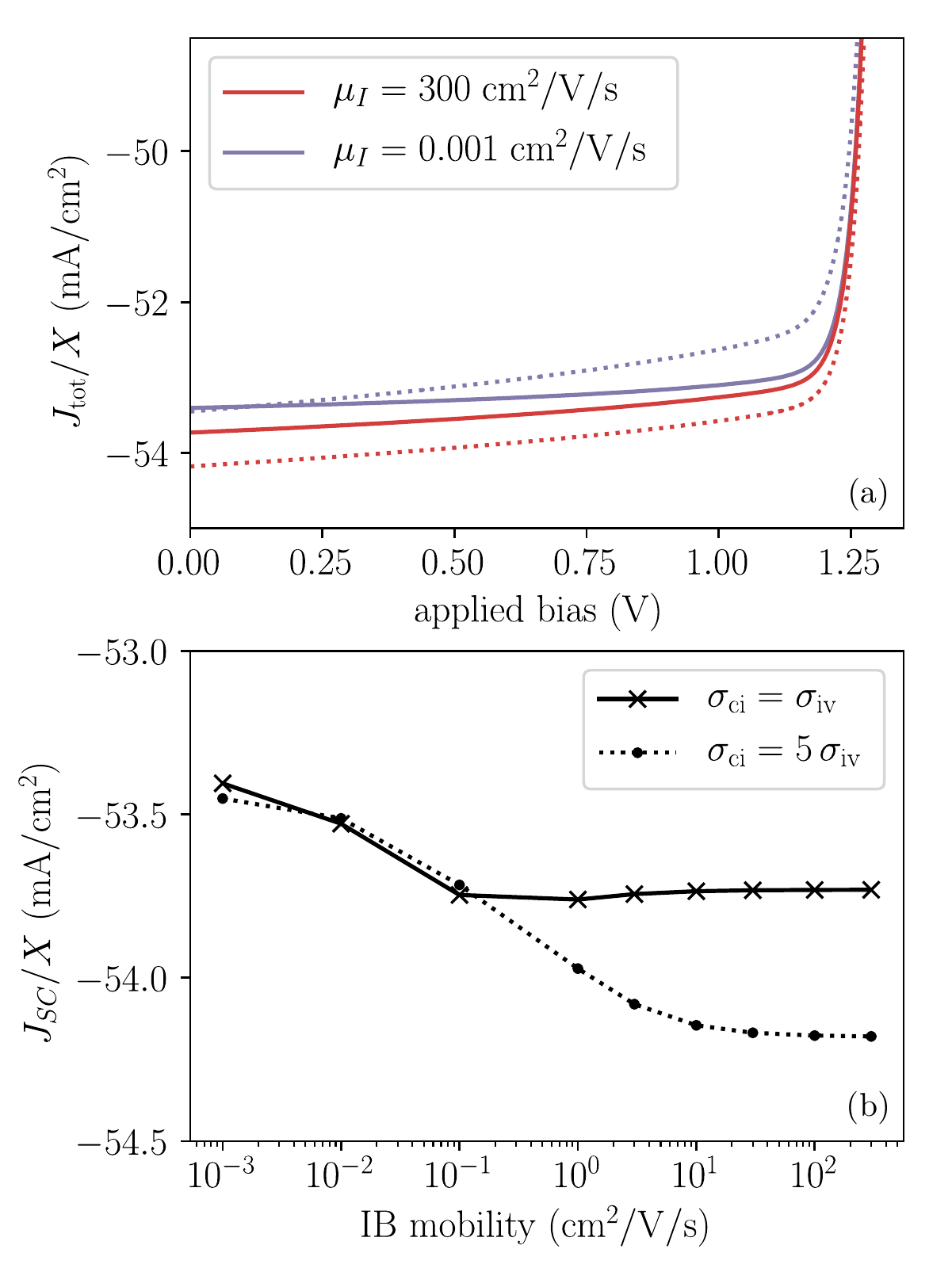}
\par\end{centering}
\caption{\label{fig:marti-JV-mis}(a) $J(V)$ curves for devices with parameters
of Table \ref{tab:marti-param}, under $X=1000$ suns illumination,
modeled on Ref.~\cite{Marti02}. Solid lines show $\sigma_{CI}^{\text{opt}}=\sigma_{IV}^{\text{opt}}$,
as in \cite{Marti02}, while dotted lines show $\sigma_{CI}^{\text{opt}}=5\sigma_{IV}^{\text{opt}}$,
which causes the CI absorption to be preferentially at the top of
the device. Note the small vertical scale. The mismatched-$\sigma^{\text{opt}}$
case is more strongly influenced by the IB mobility $\mu_{I}$, but
the effects are relatively small throughout. (b) The short-circuit
current $J_{sc}$ for devices with varying $\mu_{I}$ shows that the
matched-$\sigma^{\text{opt}}$ case is independent of $\mu_{I}$ when
$\mu_{I}$ is sufficiently large ($\gtrsim0.1$~cm$^{2}$/Vs), while
the mismatched-$\sigma^{\text{opt}}$ case again shows a stronger
$\mu_{I}$-dependence, but note the small vertical scale.}
\end{figure}

\begin{figure}
\noindent \begin{centering}
\includegraphics[width=9cm]{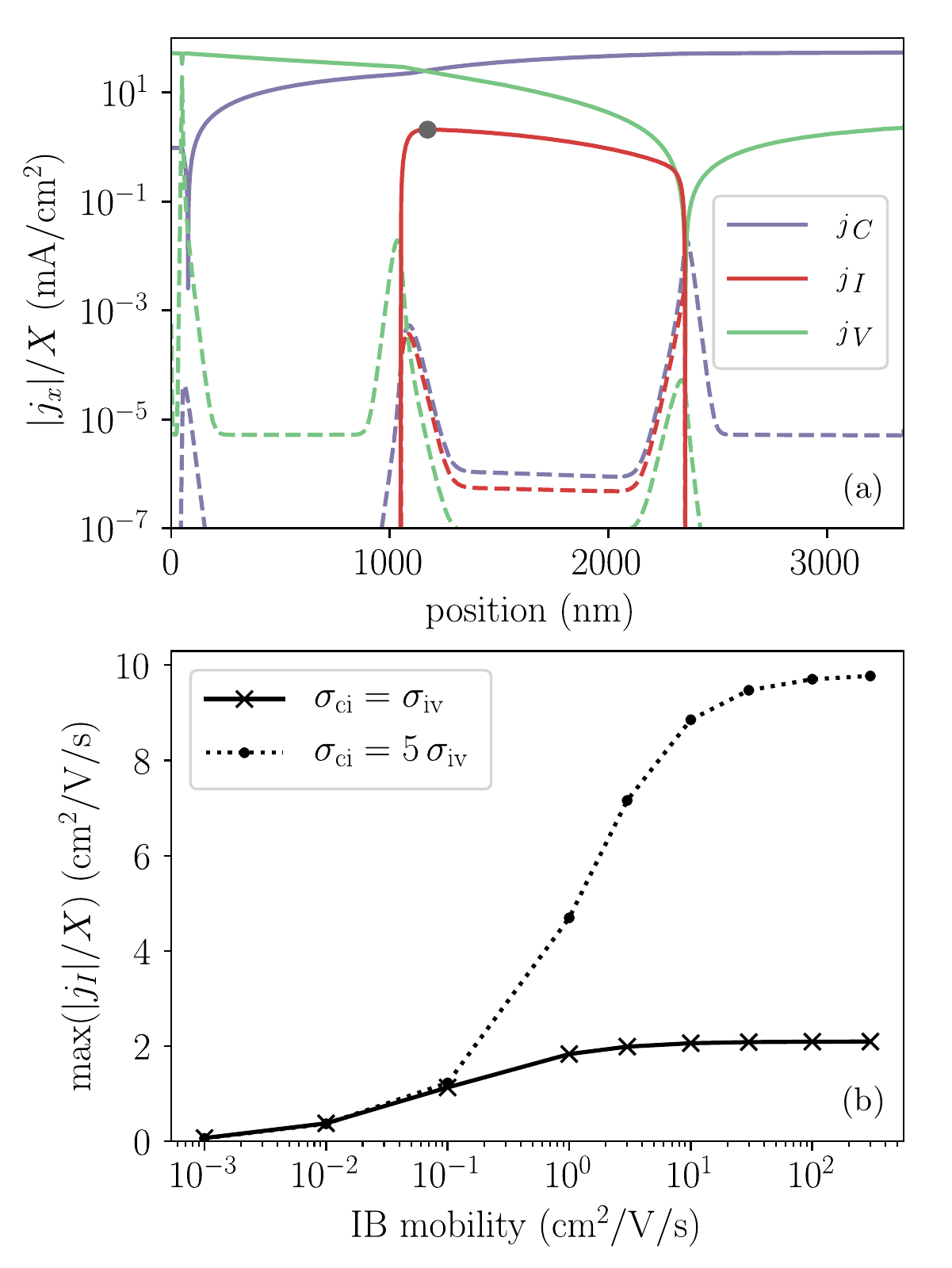}
\par\end{centering}
\caption{\label{fig:marti-j-IB-max} (a) Drift (dashed) and diffusion (solid)
currents for each band, at $\mu_{I}=100\protect\u{cm^{2}/V/s}$ and
$\sigma_{CI}^{\text{opt}}=\sigma_{IV}^{\text{opt}}$, with other parameters
as in Table \ref{tab:marti-param}. The IB current density attains
its maximum at the marked point. (b) Maximum IB current density as
a function of $\mu_{I}$ for the matched and unmatched absorption
cross sections, showing the increased importance of $j_{I}$ and thus
$\mu_{I}$ in the case where local absorptions are mismatched.}
\end{figure}

\begin{figure}
\noindent \begin{centering}
\includegraphics[width=9cm]{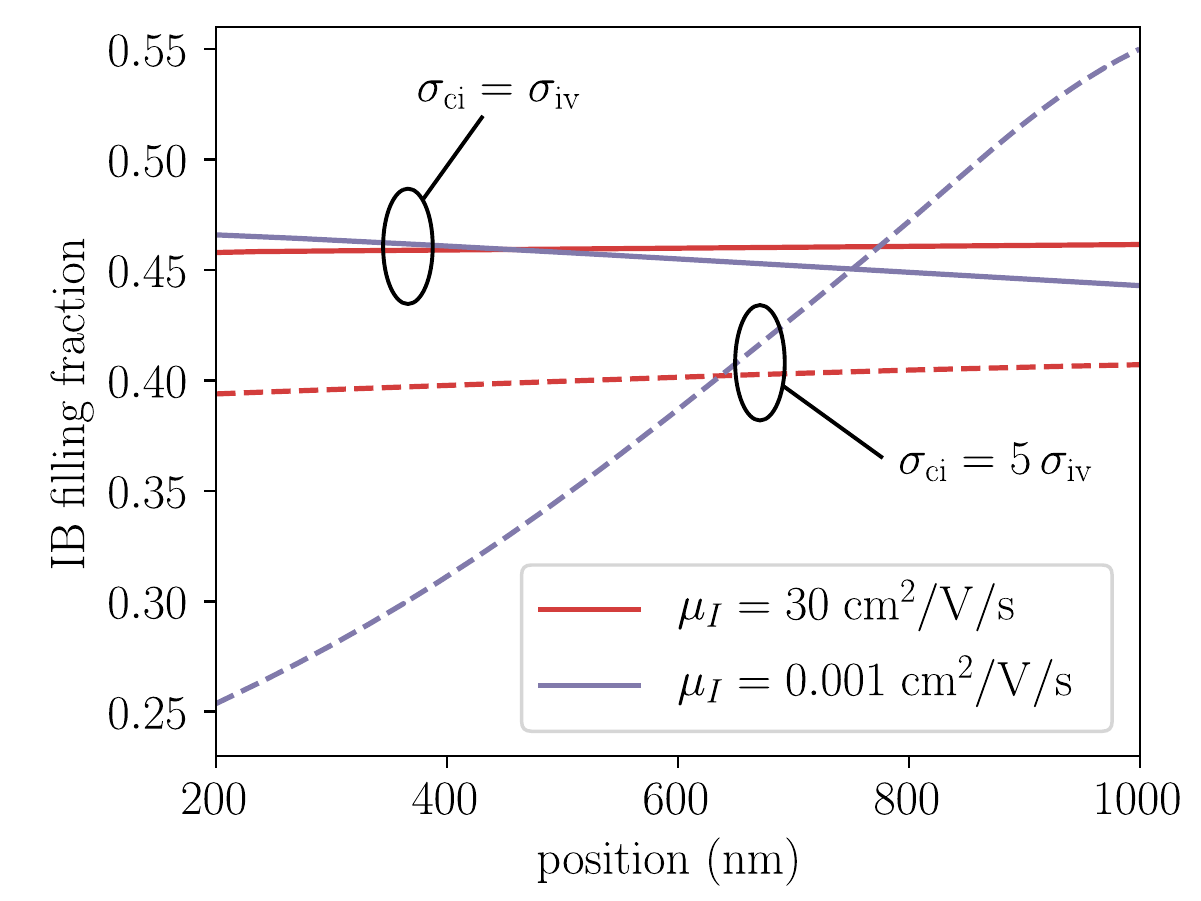}
\par\end{centering}
\caption{\label{fig:marti-f-IB} IB filling fraction $f(x)$ at the maximum
power point with the parameters as in Table \ref{tab:marti-param}.
With mismatched $\sigma^{\text{opt}}$ and low $\mu_{I}$, the IB
filling fraction changes drastically through the depth of the device,
as the local generation and recombination rates must come into balance;
this balancing increases the local recombination and decreases the
total current generation, as seen in Fig.~\ref{fig:marti-JV-mis}b.
With matched $\sigma^{\text{opt}}$, internal IB currents balance
the absorptions, and $f(x)$ remains nearly constant.}
\end{figure}

We model a similar device with a simpler p-IB-n structure. This device
has the same band and absorption parameters as the device described
in \cite{Marti2002a}, summarized in Table \ref{tab:marti-param}.
The incident light is a blackbody spectrum at $6000\u K$ with a solar
concentration factor $X=1000$. The device has equal subgap optical
cross sections and nearly current matched (within 10\%) incident photon
fluxes for the two subgap transitions. In this case, the local IV
and CI generation rates are nearly identical throughout the IB device.
The code to set up this problem is included in the supplementary material,
in $\texttt{marti2002.py}$.

We simulate this device with $\mu_{I}$ ranging from $0.001$ to $300\u{cm^{2}/V/s}$,
with resulting J-V curves shown as solid lines in Fig.~\ref{fig:marti-JV-mis}a,
which is tightly zoomed and still shows only minor effects of this
over-$10^{5}$ change in $\mu_{I}$. In fact, the IB and drifts currents
contribute negligibly to the transport inside the IB region and the
current remains diffusion-dominated throughout, as shown in Fig.~\ref{fig:marti-j-IB-max}a.

The device behavior is approximately independent of $\mu_{I}$ because
the IV and CI generation rates are roughly equal at each point in
the device, so IB transport is not required to enable sub-gap current
matching, but this behavior is not generic for all IB devices. We
illustrate this effect by increasing $\sigma_{CI}^{\textnormal{opt}}$
by a factor of five (while keeping $\sigma_{IV}^{\textnormal{opt}}$
unchanged). In this case, the device is still globally approximately
current-matched, but the CI absorption process occurs preferentially
at the top of the device while the IV generation occurs deeper in
the IB region. The device thus relies on IB transport for the CI and
IV generation rates to balance over the full device. These effects
are shown in the dashed curves of Fig.~\ref{fig:marti-JV-mis}, which
show a stronger dependence on $\mu_{I}$ than in the matched case.
When IB mobility is low, the excess CI generation in the front of
the device instead causes local CI trapping, with equivalent local
IV trapping toward the back of the device, reducing overall current.
In the high-mobility case, the overall current is slightly larger
with the mismatched absorptions, due to the increased optical depth.
Figure \ref{fig:marti-JV-mis}b shows how $J_{\text{sc}}$ varies
with $\mu_{I}$ in both of these cases, where the greater dependence
on $\mu_{I}$ in the mismatched case is apparent. Figure \ref{fig:marti-j-IB-max}b
shows that in the matched-$\sigma^{\text{opt}}$ case, $j_{I}$ is
never particularly large, while it grows to be five times larger in
the mismatched case, showing the role of IB currents in internally
balancing the optical absorptions.

In the low-mobility limit, where $j_{I}$ is always small, when local
CI and IV current generations are imbalanced, the filling fraction
$f$ of the IB must shift to equalize generation and recombination
at each point \cite{Strandberg09}. This effect is visible in Fig.~\ref{fig:marti-f-IB},
where at low mobility, the mismatched-$\sigma^{\text{opt }}$ case
has photodepletion at the front side and photofilling at the back
side, consistent with excess CI generation at the front and excess
IV generation at the back. Both the matched-$\sigma^{\text{opt }}$
and the high-mobility mismatched-$\sigma^{\textnormal{opt}}$ cases
maintain an approximately uniform IB filling fraction.

These examples together show the utility of Simudo to explore the
performance of IB devices and resolve an assertion made in earlier
device models without the benefit of a coupled PDD/optics solver.

\section{Conclusion}

The availability of a device model for intermediate band materials
should enable both understanding of this new class of materials and
optimization of IB devices. Simudo's use of the FEM and its methods
for overcoming catastrophic cancellation may also prove useful in
standard semiconductor device simulation. Simudo has been validated
against Synopsys Sentaurus for standard semiconductor devices and
shown to converge more rapidly with mesh size. This self-consistent
solution of the Poisson/drift-diffusion and optical propagation equations
provides a platform for studying a wide range of optoelectronic materials
and devices, including solar cells and photodetectors, with tools
to enable extensibility to arbitrary generation and recombination
models, thermal effects, and more. The near-term roadmap for Simudo
includes explicit heterojunction support and nonlocal tunneling, which
will be available with future releases at \href{https://github.com/simudo/simudo}{github.com/simudo/simudo}.
We hope that the free and open source nature of this software will
enable further development of IB materials and device simulation more
broadly.

\section*{Acknowledgments}

We acknowledge funding from US Army Research Laboratory (W911NF-16-2-0167),
the Natural Sciences and Engineering Research Council of Canada TOP-SET
training program, and computing resources from Compute Canada. We
thank Emily Zinnia Zhang for alpha testing Simudo, contributing the
first code implementing trapping processes, and valuable conversations.

\bibliographystyle{spphys}


\begin{thebibliography}{10}
\providecommand{\url}[1]{{#1}}
\providecommand{\urlprefix}{URL }
\expandafter\ifx\csname urlstyle\endcsname\relax
  \providecommand{\doi}[1]{DOI \discretionary{}{}{}#1}\else
  \providecommand{\doi}{DOI \discretionary{}{}{}\begingroup
  \urlstyle{rm}\Url}\fi

\bibitem{Bank1983}
R.E. Bank, D.J. Rose, W.~Fichtner, \emph{{Numerical Methods for Semiconductor
  Device Simulation}}, IEEE Transactions on Electron Devices \textbf{30}(9),
  1031 (1983)

\bibitem{Fichtner1983a}
W.~Fichtner, D.~Rose, R.~Bank, \emph{{Semiconductor device simulation}}, IEEE
  Transactions on Electron Devices \textbf{30}(9), 1018 (1983)

\bibitem{Markowich1986}
P.A. Markowich, \emph{{The Stationary Semiconductor Device Equations}}.
\newblock Computational Microelectronics (Springer Vienna, Vienna, 1986)

\bibitem{Piprek18}
J.~Piprek (ed.), \emph{{Handbook of Optoelectronic Device Modeling {\&}
  Simulation}} (CRC Press, Boca Raton, FL, 2018)

\bibitem{Schenk1998}
A.~Schenk, \emph{{Advanced Physical Models for Silicon Device Simulation}}
  (Springer-Verlag Wien, 1998)

\bibitem{Luque97}
A.~Luque, A.~Mart\'i, \emph{Increasing the Efficiency of Ideal Solar Cells by
  Photon Induced Transitions at Intermediate Levels}, Phys. Rev. Lett.
  \textbf{78}(26), 5014 (1997)

\bibitem{Okada15}
Y.~Okada, N.J. Ekins-Daukes, T.~Kita, R.~Tamaki, M.~Yoshida, A.~Pusch, O.~Hess,
  C.C. Phillips, D.J. Farrell, K.~Yoshida, N.~Ahsan, Y.~Shoji, T.~Sogabe, J.F.
  Guillemoles, \emph{Intermediate band solar cells: Recent progress and future
  directions}, Applied Physics Reviews \textbf{2}(2), 021302 (2015)

\bibitem{Mailoa14}
J.P. Mailoa, A.J. Akey, C.B. Simmons, D.~Hutchinson, J.~Mathews, J.T. Sullivan,
  D.~Recht, M.T. Winkler, J.S. Williams, J.M. Warrender, P.D. Persans, M.J.
  Aziz, T.~Buonassisi, \emph{Room-temperature sub-band gap optoelectronic
  response of hyperdoped silicon}, Nat Commun \textbf{5}, 3011 (2014)

\bibitem{Berencen17}
Y.~Berenc{\'{e}}n, S.~Prucnal, F.~Liu, I.~Skorupa, R.~H\"ubner, L.~Rebohle,
  S.~Zhou, H.~Schneider, M.~Helm, W.~Skorupa, \emph{Room-temperature
  short-wavelength infrared {Si} photodetector}, Scientific Reports \textbf{7},
  43688 (2017)

\bibitem{Wang18}
M.~Wang, Y.~Berenc\'en, E.~Garc\'ia-Hemme, S.~Prucnal, R.~H\"ubner, Y.~Yuan,
  C.~Xu, L.~Rebohle, R.~B\"ottger, R.~Heller, H.~Schneider, W.~Skorupa,
  M.~Helm, S.~Zhou, \emph{Extended Infrared Photoresponse in
  $\mathrm{Te}$-Hyperdoped $\mathrm{Si}$ at Room Temperature}, Phys. Rev.
  Applied \textbf{10}(2), 024054 (2018)

\bibitem{Brown02}
A.S. Brown, M.A. Green, \emph{Impurity photovoltaic effect: Fundamental energy
  conversion efficiency limits}, J. Appl. Phys. \textbf{92}(3), 1329 (2002)

\bibitem{Marti06}
A.~Mart\'i, E.~Antol\'in, C.R. Stanley, C.D. Farmer, N.~L\'opez, P.~D\'iaz,
  E.~C\'anovas, P.G. Linares, A.~Luque, \emph{Production of Photocurrent due to
  Intermediate-to-Conduction-Band Transitions: A Demonstration of a Key
  Operating Principle of the Intermediate-Band Solar Cell}, Phys. Rev. Lett.
  \textbf{97}(24), 247701 (2006)

\bibitem{Wang09}
W.~Wang, A.S. Lin, J.D. Phillips, \emph{Intermediate-band photovoltaic solar
  cell based on {ZnTe:O}}, Appl. Phys. Lett. \textbf{95}(1), 011103 (2009)

\bibitem{Lopez11}
N.~L\'opez, L.A. Reichertz, K.M. Yu, K.~Campman, W.~Walukiewicz,
  \emph{Engineering the Electronic Band Structure for Multiband Solar Cells},
  Phys. Rev. Lett. \textbf{106}(2), 028701 (2011)

\bibitem{Sullivan15}
J.T. Sullivan, C.B. Simmons, T.~Buonassisi, J.J. Krich, \emph{Targeted Search
  for Effective Intermediate Band Solar Cell Materials}, IEEE Journal of
  Photovoltaics \textbf{5}(1), 212 (2015)

\bibitem{Maur08}
M.A. der Maur, A multiscale simulation environment for electronic and
  optoelectronic devices.
\newblock Ph.D. thesis, Universita' degli Studi di Roma Tor Vergata (2008)

\bibitem{Birner2007}
S.~Birner, T.~Zibold, T.~Andlauer, T.~Kubis, M.~Sabathil, A.~Trellakis,
  P.~Vogl, \emph{{nextnano: General Purpose 3-D Simulations}}, IEEE
  Transactions on Electron Devices \textbf{54}(9), 2137 (2007)

\bibitem{Clugston97}
D.A. Clugston, P.A. Basore, \emph{{PC1D} version 5: 32-bit solar cell modeling
  on personal computers}, in \emph{Twenty Sixth IEEE Photovoltaic Specialists
  Conference} (1997), pp. 207--210

\bibitem{Haug16}
H.~Haug, J.~Greulich, \emph{{PC1Dmod 6.2} - Improved Simulation of {c-Si}
  Devices with Updates on Device Physics and User Interface}, Energy Procedia
  \textbf{92}(1876), 60 (2016)

\bibitem{Varache15}
R.~Varache, C.~Leendertz, M.~Gueunier-Farret, J.~Haschke, D.~Mu\~noz, L.~Korte,
  \emph{Investigation of selective junctions using a newly developed tunnel
  current model for solar cell applications}, Solar Energy Materials and Solar
  Cells \textbf{141}, 14 (2015)

\bibitem{Burgelman00}
M.~Burgelman, P.~Nollet, S.~Degrave, \emph{Modelling polycrystalline
  semiconductor solar cells}, Thin Solid Films \textbf{361-362}, 527 (2000)

\bibitem{Alonso-Alvarez2018}
D.~Alonso-{\'{A}}lvarez, T.~Wilson, P.~Pearce, M.~F{\"{u}}hrer, D.~Farrell,
  N.~Ekins-Daukes, \emph{{Solcore: a multi-scale, Python-based library for
  modelling solar cells and semiconductor materials}}, Journal of Computational
  Electronics \textbf{17}(3), 1099 (2018)

\bibitem{Shockley52}
W.~Shockley, W.T. Read, \emph{Statistics of the Recombinations of Holes and
  Electrons}, Phys. Rev. \textbf{87}(5), 835 (1952)

\bibitem{Marti02}
A.~Marti, L.~Cuadra, A.~Luque, \emph{Quasi-drift diffusion model for the
  quantum dot intermediate band solar cell}, IEEE Transactions on Electron
  Devices \textbf{49}(9), 1632 (2002)

\bibitem{Strandberg11}
R.~Strandberg, T.W. Reenaas, \emph{Drift-diffusion model for intermediate band
  solar cells including photofilling effects}, Prog. Photovolt: Res. Appl.
  \textbf{19}(1), 21 (2011)

\bibitem{Tobias11}
I.~Tob\'ias, A.~Luque, A.~Mart\'i, \emph{Numerical modeling of intermediate
  band solar cells}, Semiconductor Science and Technology \textbf{26}(1),
  014031 (2011)

\bibitem{Yoshida12a}
K.~Yoshida, Y.~Okada, N.~Sano, \emph{Device simulation of intermediate band
  solar cells: Effects of doping and concentration}, Journal of Applied Physics
  \textbf{112}(8),  (2012)

\bibitem{Cuadra04a}
L.~Cuadra, A.~Mart\'i, A.~Luque, \emph{Influence of the overlap between the
  absorption coefficients on the efficiency of the intermediate band solar
  cell}, IEEE Transactions on Electron Devices \textbf{51}(6), 1002 (2004)

\bibitem{Levy08}
M.Y. Levy, C.~Honsberg, \emph{Intraband absorption in solar cells with an
  intermediate band}, Journal of Applied Physics \textbf{104}(11), 113103
  (2008)

\bibitem{Hu10}
W.G. Hu, T.~Inoue, O.~Kojima, T.~Kita, \emph{Effects of absorption coefficients
  and intermediate-band filling in InAs/GaAs quantum dot solar cells}, Appl.
  Phys. Lett. \textbf{97}(19), 193106 (2010)

\bibitem{Strandberg17}
R.~Strandberg, \emph{Analytic ${JV}$-Characteristics of Ideal Intermediate Band
  Solar Cells and Solar Cells With Up and Downconverters}, IEEE Transactions on
  Electron Devices \textbf{64}(5), 2275 (2017)

\bibitem{Strandberg17a}
R.~Strandberg, \emph{The ${JV}$-Characteristic of Intermediate Band Solar Cells
  With Overlapping Absorption Coefficients}, IEEE Transactions on Electron
  Devices \textbf{64}(12), 5027 (2017)

\bibitem{Lin09}
A.S. Lin, W.~Wang, J.D. Phillips, \emph{Model for intermediate band solar cells
  incorporating carrier transport and recombination}, Journal of Applied
  Physics \textbf{105}(6), 064512 (2009)

\bibitem{Navruz08}
T.~Navruz, M.~Saritas, \emph{Efficiency variation of the intermediate band
  solar cell due to the overlap between absorption coefficients}, Solar Energy
  Materials and Solar Cells \textbf{92}(3), 273 (2008)

\bibitem{Krich14}
J.J. Krich, A.H. Trojnar, L.~Feng, K.~Hinzer, A.W. Walker, \emph{Modeling
  intermediate band solar cells: a roadmap to high efficiency}, in \emph{Proc.
  SPIE 8981, Physics, Simulation, and Photonic Engineering of Photovoltaic
  Devices III} (2014), p. 89810O

\bibitem{Yoshida10}
K.~Yoshida, Y.~Okada, N.~Sano, \emph{Self-consistent simulation of intermediate
  band solar cells: Effect of occupation rates on device characteristics},
  Applied Physics Letters \textbf{97}(13), 133503 (2010)

\bibitem{AlnaesBlechta2015a}
M.S. Aln{\ae}s, J.~Blechta, J.~Hake, A.~Johansson, B.~Kehlet, A.~Logg,
  C.~Richardson, J.~Ring, M.E. Rognes, G.N. Wells, \emph{The FEniCS Project
  Version 1.5}, Archive of Numerical Software \textbf{3}(100) (2015)

\bibitem{Eymard00}
R.~Eymard, T.~Gallou\"et, R.~Herbin, in \emph{Handbook of Numerical Analysis},
  vol.~7 (Elsevier, 2000), pp. 713--1018

\bibitem{He91}
Y.~He, G.~Cao, \emph{A generalized Scharfetter-Gummel method to eliminate
  crosswind effects (semiconduction device modeling)}, IEEE Transactions on
  Computer-Aided Design of Integrated Circuits and Systems \textbf{10}(12),
  1579 (1991)

\bibitem{Nachaoui99}
A.~Nachaoui, \emph{{Iterative solution of the drift-diffusion equations}},
  Numerical Algorithms \textbf{21}, 323 (1999)

\bibitem{Bochev15}
P.~Bochev, K.~Peterson, M.~Perego, \emph{A multiscale control volume finite
  element method for advection-diffusion equations}, Int. J. Numer. Meth.
  Fluids \textbf{77}(11), 641 (2015)

\bibitem{Cockburn00}
B.~Cockburn, G.E. Karniadakis, C.W. Shu (eds.).
\newblock \emph{Discontinuous Galerkin Methods: Theory, Computation, and
  Applications}, \emph{Lecture Notes in Computational Science and Engineering},
  vol.~11 (Springer, Berlin, Heidelberg, 2000)

\bibitem{Kumar16}
G.~Kumar, M.~Singh, A.~Bulusu, G.~Trivedi, \emph{A Framework to Simulate
  Semiconductor Devices Using Parallel Computer Architecture}, in \emph{Journal
  of Physics: Conference Series}, vol. 759 (IOP Publishing, 2016), vol. 759,
  pp. 012,098--

\bibitem{Kumar17}
G.~Kumar, M.~Singh, A.~Ray, G.~Trivedi, \emph{An FEM based framework to
  simulate semiconductor devices using streamline upwind Petrov-Galerkin
  stabilization technique}, in \emph{2017 27th International Conference
  Radioelektronika} (2017), pp. 1--5

\bibitem{Poupaud1991a}
F.~Poupaud, C.~Schmeiser, \emph{Charge transport in semiconductors with
  degeneracy effects}, Mathematical methods in the applied sciences
  \textbf{14}(5), 301 (1991)

\bibitem{Marshak1984a}
A.H. Marshak, C.~Van~Vliet, \emph{Electrical current and carrier density in
  degenerate materials with nonuniform band structure}, Proceedings of the IEEE
  \textbf{72}(2), 148 (1984)

\bibitem{Nelson03}
J.~Nelson, \emph{The Physics of Solar Cells} (Imperial College Press, 2003)

\bibitem{McIntosh14}
K.R. McIntosh, L.E. Black, \emph{{On effective surface recombination
  parameters}}, Journal of Applied Physics \textbf{116}(1) (2014)

\bibitem{Zhao2013a}
J.~Zhao, J.~Tan, L.~Liu, \emph{A second order radiative transfer equation and
  its solution by meshless method with application to strongly inhomogeneous
  media}, Journal of Computational Physics \textbf{232}(1), 431  (2013)

\bibitem{Johnson1987a}
C.~Johnson, \emph{Numerical solution of partial differential equations by the
  finite element method} (Cambridge University Press, 1987)

\bibitem{Gockenbach2006a}
M.S. Gockenbach, \emph{Understanding and implementing the finite element
  method} (SIAM, 2006)

\bibitem{Donea2003a}
A.H. Jean~Donea, \emph{Finite element methods for flow problems} (Wiley, 2003)

\bibitem{Gaury2019a}
B.~{Gaury}, Y.~{Sun}, P.~{Bermel}, P.M. {Haney}, \emph{Sesame: a 2-dimensional
  solar cell modeling tool}, Solar Energy Materials and Solar Cells
  \textbf{198}, 53 (2019)

\bibitem{Brezzi1985a}
F.~Brezzi, J.~Douglas, L.D. Marini, \emph{Two families of mixed finite elements
  for second order elliptic problems}, Numerische Mathematik \textbf{47}(2),
  217 (1985)

\bibitem{Roberts1991a}
J.~Roberts, J.M. Thomas, in \emph{Finite Element Methods (Part 1)},
  \emph{Handbook of Numerical Analysis}, vol.~2 (Elsevier, 1991), pp. 523--639

\bibitem{Cummings09}
D.J. Cummings, M.E. Law, S.~Cea, T.~Linton, \emph{{Comparison of discretization
  methods for device simulation}}, International Conference on Simulation of
  Semiconductor Processes and Devices 2009 pp. 1--4 (2009)

\bibitem{Synopsys15}
{Synopsys Inc.}, \emph{{Sentaurus Device User Guide, vK-2015}} (Synopsys Inc.,
  2015)

\bibitem{LoggMardalEtAl2012a}
A.~Logg, K.A. Mardal, G.N. Wells, et~al., \emph{Automated Solution of
  Differential Equations by the Finite Element Method} (Springer, 2012)

\bibitem{Marti2002a}
A.~{Marti}, L.~{Cuadra}, A.~{Luque}, \emph{{Quasi-drift diffusion model for the
  quantum dot intermediate band solar cell}}, IEEE Transactions on Electron
  Devices \textbf{49}, 1632 (2002)

\bibitem{Strandberg09}
R.~Strandberg, T.W. Reenaas, \emph{Photofilling of intermediate bands}, J.
  Appl. Phys. \textbf{105}(12), 124512 (2009)

\end{thebibliography}

\end{document}